\documentclass[12pt]{article}
\usepackage{ifpdf}
\pdfoutput=1
\usepackage{a4wide}
\usepackage{amsmath,amsthm,amssymb}
\usepackage{graphics,graphicx}
\usepackage{epsfig}
\usepackage{color}

\begin{document}

\rightline{PUPT-2262}
\rightline{QMUL-PH-08-06}
\vspace{2truecm}

\centerline{\LARGE \bf Electromagnetic form factors }

\vspace{0.5cm}

\centerline{\LARGE \bf from the fifth dimension}

\vspace{1.3truecm}

\centerline{
    {\large \bf D. Rodr\'{\i}guez-G\'omez${}^{a,b,}$}\footnote{drodrigu@princeton.edu}
    {\bf and}
    {\large \bf J. Ward${}^{c,}$}\footnote{jwa@uvic.ca}}

\vspace{.4cm}
\centerline{{\it ${}^a$Department of Physics, 
Princeton University}}
\centerline{{\it Princeton, NJ 08544, USA}}

\vspace{.4cm}
\centerline{{\it ${}^b$ Center for Research in String Theory, Queen Mary University of
    London}} \centerline{{\it Mile End Road, London, E1 4NS, UK}}

\vspace{.4cm}
\centerline{{\it ${}^c$ Department of Physics and Astronomy, University of
    Victoria}} \centerline{{\it Victoria, BC, V8P 1A1, Canada}}

\vspace{2truecm}

\centerline{\bf ABSTRACT}
\vspace{.5truecm}

\noindent
We analyse various $U(1)_{EM}$ form factors of mesons at strong coupling in an $\mathcal{N}=2$ flavored version of $\mathcal{N}=4$ $SYM$ which becomes conformal in the UV. 
The quark mass breaks the conformal symmetry in the IR and generates a mass gap. In the appropriate limit, the gravity dual is described in terms of 
probe $D7$-branes in $AdS_5\times S^5$. By studying the $D7$ fluctuations we find the suitable terms in a ``meson effective theory" which allow us to compute the 
desired form factors, namely the $\gamma\pi\rho$ and $\gamma f_0\rho$ transition form factors. At large $q^2$ we find perfect agreement with the naive parton model counting, which is a consequence of the conformal nature of both QCD and 
our model in the UV. By using the same tools, we can compute the $\gamma^*\gamma^*\pi$ form factor. However this channel is more subtle and comparisons to the QCD result are more involved.  

\newpage
\tableofcontents
\section{Introduction}

Understanding the generic behavior of gauge theories remains as one of the most fundamental problems in theoretical physics. 
At weak coupling a perturbative treatment is amenable, however the strong coupling dynamics represents an incredible challenge. 
It is believed that this regime can be understood in terms of a string theory. 
This correspondence has been made more precise for a certain class of gauge theories over the past decade, through the use of gauge/gravity duality \cite{Maldacena:1997re}. 

It is known that the dynamics of gauge theories differs significantly depending on whether or not they contain fields in the fundamental representation of the gauge group. 
One of the most obvious features of having such fields is that there is the possibility of forming bound states. 
At weak coupling these bound states appear as positronium, \textit{i.e.} a system analogous to the hydrogen atom but composed of a quark and an antiquark. 
In order to probe the strong coupling dynamics of flavored gauge theories, it is interesting to study these objects at large $\lambda$, where $\lambda$ is the 't Hooft 
coupling. A very natural tool to adopt is that of gauge/gravity duality. 
However including fundamental matter is a difficult problem. 
A step forward was taken in \cite{Karch:2000ct, Karch:2000gx, Karch:2002sh, Karch:2002xe}, were it was suggested to introduce flavor as a new open string sector 
coming from an extra stack of branes (so-called 'flavor' branes) intersecting the color branes. In the limit in which the number of flavor branes is so 
small that they can be considered a small perturbation, we can perform the geometric transition and replace the color branes by their near horizon 
geometry - where we should consider the flavor branes as probes. 
This sort of quenched approximation has a number of consequences, one of which is that the running of quarks in loops is absent. 
This translates, in particular, into a vanishing beta function for the gauge coupling. However, in the case of massive flavors, the conformal symmetry is 
broken in the IR, leading to the existence of ``mesons". These bound states were studied for the first time in \cite{Kruczenski:2003be} 
(for reviews see \cite{RodriguezGomez:2007za, Erdmenger:2007cm}).  
It is only very recently that fully backreacted solutions, corresponding to an unquenched approach, have been found in \cite{Casero:2006pt, Casero:2007jj, Benini:2006hh, Benini:2007gx, Benini:2007kg, Canoura:2008at}.

In this paper we will be interested in the strong coupling structure of these mesons. Following the approach in  \cite{Hong:2003jm}, we will probe them with photons. As anticipated, in the approximation we will work on, the beta function for the gauge coupling vanishes. Then, it is to be expected the large momentum transfer regime of the scatterings we wil be computing, which is insensitive to the IR relevant mass term, to be controlled by conformal invariance. Related processes have been considered in the literature using a gravity dual for QCD, such as \cite{  Brodsky:2003px, deTeramond:2005su, Brodsky:2006uqa, Brodsky:2007hb, Brodsky:2008pg, BallonBayona:2007rs, BallonBayona:2007qr, BoschiFilho:2002vd, BoschiFilho:2002ta}, and also \cite{Brodsky:2008pf, Abidin:2008hn} where gravitational form factors have been computed. Note that in our case, the gravity dual captures the strong coupling regime of the theory. Thus, as opposed to real QCD, in our case the large momentum transfer regime will be dominated by a strongly coupled conformal theory. The fact that conformal invariance is recovered in the UV is translated into an appropriate dictionary which allows to use the scaling coming from naive parton counting valid at weak coupling, along the lines of \cite{Polchinski:2001tt,Polchinski:2002jw,Hong:2003jm}. 

In order to study the mesons, we will consider the 
simplest theory admitting a gravity dual and containing a mass gap, which can be engineered as a $D3$-$D7$ intersection in flat space.\footnote{The massless limit of this theory was considered in \cite{McGreevy:2007kt}, where quark scattering is computed along the lines of \cite{Alday:2007hr}. It would be interesting to apply these methods to the massive (non-conformal) case.} 
In more adequate terminology, we will be computing 
electromagnetic transition form factors. This requires to couple the gauge theory to electromagnetism, however from the point of view of the $SU(N_c)$ 
dynamics, the $U(1)_{EM}$ is just a global symmetry. Technically this allows us to consider the EM current as a $U(1)$ subgroup in the $SU(N_f)$, 
corresponding to the gauge field on the flavor brane.\footnote{For example, this approach is similar to that in \cite{Mateos:2007yp}.} This will require us to find the adequate couplings in the meson effective theory, allowing us to compute the desired form factors. Note that in \cite{Hong:2003jm} the vector field probing the mesons was the full $SU(N_f)$. After introducing the field theory and its gravity dual in section 2, we derive the corresponding interaction lagrangian allowing us to compute such form factors in 
section 3. In section 4 we compute and analyze these transition form factors. 
In accordance with the results in \cite{Hong:2003jm}, we are able to match the expectations from QCD at large momentum transfer. 
This is to be expected since, in that regime, both QCD and our theory are dominated by conformal invariance. 
Interestingly we can make use of the interaction lagrangian to compute the process $\gamma^*\gamma^*\pi$. As opposed to the form factors, this case is more contrived and we do not have a fully satisfactory field theory picture. On the other hand, this process will be related to the $\gamma^*\pi\rho$ form factor due to vector meson dominance in much the same spirit as in QCD. 
In section 5 we examine the full amplitude, in which the analog of the hadronic tensor exhibits a Callan-Gross relation. 
This is deeply connected with the helicity structure of our amplitudes. We finish in section 6 with some comments and suggestions for future directions.

\section{The field theory and its gravity dual}

The theory in question consists of $\mathcal{N}=4$ SYM coupled to $N_f$ fundamental hypermultiplets in such a way that the 
final theory preserves $\mathcal{N}=2$ supersymmetry. 
Generically the hypermultiplets will be massive, and we will assume a diagonal mass matrix. 
It is important to note that our theory is non-chiral even in the massless limit. In particular this means that the flavor symmetry is just $SU(N_f)$. 
The field content is

\begin{center}
\begin{tabular}{|c|c|c|}
\hline
 & $SU(N_c)$ & $SU(N_f)$\\ \hline
$\Phi_I$ & $Adj$ &\textbf{1} \\ \hline
$Q^i$ & $\Box$ & $\bar{\Box}$\\ \hline
$\tilde{Q}_i$ & $\bar{\Box}$ & $\Box$\\ \hline
\end{tabular}
\end{center}
and the superpotential reads

\begin{equation}
\label{W}
W=\tilde{Q}_i(m_q+\Phi_3)Q^i+\Phi_I\Phi_J\Phi_K\epsilon^{IJK}\ ,
\end{equation}
The $\Phi_I,\, I=1,2,3$ are the 3 chiral superfields of the $\mathcal{N}=4$ SYM sector, whilst the $(Q,\tilde{Q})$ flavor hypermultiplets 
break the supersymmetry down to $\mathcal{N}=2$.
The mass term additionally breaks the $U(1)_R$ symmetry. Let us set $m_q$ to zero for a moment. 
In that case there is an $R$-symmetry under which $R_{Q}=R_{\Phi_I}=\frac{2}{3}$.
\footnote{This R-charge assignation is the one coming from a-maximization, and is indeed the 
one adapted to match the beta function coming from the gravity description (see for example \cite{Franco:2004jz}, or \cite{Nification} for a discussion with D7 branes).}
Assuming we are close to a conformal fixed point, 
we can compute the exact beta function of the theory by approximating $\gamma_i\sim 3R_i-2$. It is then straightforward to see that

\begin{equation}
\beta_{g_{YM}}=\frac{d}{d\log\mu} \frac{8\pi^2}{g_{YM}^2}=-N_f\ .
\end{equation}
Thus we see that the theory is not asymptotically free, but rather develops a Landau pole in the UV. However we will treat the theory in the large $N_c$ limit. 
The beta function for the 't Hooft coupling reads

\begin{equation}
\beta_{\lambda}=\frac{d}{d\log\mu} \frac{8\pi^2}{g_{YM}^2N_c}=\frac{d}{d\log\mu} \frac{8\pi^2}{\lambda}=-\frac{N_f}{N_c}\ .
\end{equation}
Therefore in the limit in which we have a large number of colors and a finite number of flavors in such a way that $N_f/N_c\sim 0$ we can still make sense of the theory since it becomes conformal. 
In the generic case in which there is a mass term, since it has a classical negative beta function, the 
UV properties will not be changed from those of the massless case, and we expect that our theory approaches a UV conformal point provided 
we are in the limit $N_f/N_c\sim 0$. Note that even though the gauge coupling has a vanishing beta function, conformal invariance will be broken in the IR by the 
scale set by $m_q$. 

It is important to note that in the $N_f/N_c\sim 0$ limit, non-perturbative effects will be negligible. For example, the ADS superpotential 
gives no contribution, since the exponent of the meson matrix is zero. 

Let us finally discuss the global symmetries. The theory has an $SU(2)_R\times SU(2)$ global symmetry, of which the $SU(2)_R$ is an 
R-symmetry (and therefore does not commute with the supercharges), whilst the other $SU(2)$ is a global symmetry. 
As noted above in the case of massless hypermultiplets the R-symmetry is enhanced back to $SU(2)_R\times U(1)_R$.

\subsection{The gravity dual}

The theory above can be engineered as a brane web. Consider the $D3$-$D7$ intersection according to the following array:
 
 \begin{equation}
\begin{array}{ccccccccccl}
  &1&2&3& 4& 5&6 &7&8&9 & \nonumber \\
N_c\, D3: & \times &\times &\times &\_ &\_ & \_&\_ &\_ &\_ &     \nonumber \\
N_f\, D7: &\times&\times&\times&\times&\times&\times&\times&\_&\_&
\end{array}
\end{equation}
Working at small 't Hooft coupling, upon taking the decoupling limit, the local dynamics on the $D7$-branes decouples and appears as a global flavor 
symmetry in the effective 4-dimensional field theory description. This field theory is precisely the one introduced above. 
The $3$-$3$ strings give rise to the $\mathcal{N}=4$ SYM fields, while the $3$-$7$ strings generate the flavor hypermultiplets. \footnote{Note that the $7$-$7$ strings are non
dynamical in the gauge theory.}

Without loss of generality let us localise the $D3$-branes at the origin of the $(8,9)$ plane. Then the $i$-th such $D7$-brane will sit at a point 
$\vec{z}^i=(x_8 ^i,x_9^i)$, which is at a distance $L^i=\sqrt{(x_8^i)^2+(x_9^i)^2}$. This distance, in units of  $2\pi\alpha'$, defines the mass of the 
$i$-th hypermultiplet. However, for simplicity, we will assume that all the masses are equal, corresponding to a configuration where all the $D7$ 
are in a single stack located at $\vec{z}=(x_8,x_9)$. In that case we recover the full $SU(N_f)$ flavor symmetry with $m_q=L / (2\pi\alpha')$.

We can provide a closed string description of the system which captures the strong coupling regime of the theory by 
considering the gravity dual of the above system. In the $N_c/N_f\sim 0$ limit the backreaction of the $D7$-branes is negligible. 
Then, an accurate description of the system can be achieved by considering $N_f$ probe $D7$ in the near horizon of the background sourced by the $N_c$ $D3$-branes. The background is then simply $AdS_5\times S^5$ which has a constant dilaton, translating into a vanishing beta function for the 
gauge theory 't Hooft coupling in agreement with our discussion above. 

In order to describe the flavor D7 embeddings, we can write the $AdS_5\times S^5$ metric as

\begin{equation}
ds^2=\frac{\vec{x}^2+\vec{z}^2}{R^2}dx_{1,3}^2+\frac{R^2}{\vec{x}^2+\vec{z}^2}(d\vec{x}^2+d\vec{z}^2)\, ,
\end{equation}
where $\vec{x}=(x^4,\cdots,x^7)$. Working in static gauge, the $D7$-branes will have as worldvolume coordinates $(x_{1,3},\,\vec{x})$, 
whilst sitting at fixed $\vec{z}^2=L^2$. It is now straightforward to write the induced metric on them in polar coordinates as
 
 \begin{equation}
ds^2_{D7}=\frac{(r^2+L^2)}{R^2} dx_{1,3}^2+\frac{R^2}{(r^2+L^2)}(dr^2+r^2d\Omega_3^2)\, ,
\end{equation}
As usual, the radial coordinate on the $D7$ will have the interpretation of holographic energy. We can now see how at large $r$, corresponding to the UV of the field theory, 
the metric approaches $AdS_5\times S^3$. Additionally, since the 't Hooft coupling is constant, we see that the theory approaches a conformal fixed point in the UV. 
However in the IR, the metric above deviates from pure $AdS$ because of the presence of the IR scale $L$. Since $m_q=L/(2\pi\alpha')$ we see that conformal invariance is 
lost because of the scale $m_q$, which introduces a mass gap in accordance with the field theory analysis above.

From the supergravity we can also read off the resulting $R$-symmetry of the field theory. The $\vec{x}$ coordinates on the $D7$ 
enjoy an $SO(4)\sim SU(2)\times SU(2)$ symmetry. However the $AdS_5\times S^5$ background also has a 4-form RR potential which can couple to the $D7$. 
Indeed the symmetry which interchanges the two $SU(2)$ is broken by the Chern-Simons term on the $D7$-branes. Therefore one of the $SU(2)$ becomes the $SU(2)_R$ 
while the other remains as the global symmetry $SU(2)$. In the case of massless quarks, the $D7$ sit on top of the $D3$-branes and therefore we recover 
rotational invariance in the $(8,9)$ plane, which corresponds to the $U(1)_R$. 

\section{Effective meson theory from SUGRA}

Since our theory is not conformal in the IR we expect it develops a mass gap, generating a meson spectrum. At weak coupling these mesons are 
positronium-like systems, however we are interested in their strong coupling description. In order to investigate this 
we should analyze the $3$-$7$ strings corresponding to the 
quark fields, but in the dual gravity description which captures the strong coupling. As we have argued before, after the geometric transition the 
strong coupling gravity dual is in terms of $N_f$ probe $D7$ in the near-horizon of the $D3$ background. 
Quarks, therefore, correspond to strings hanging from the flavor branes, and quark bound states, \textit{i.e.} the mesons we want to study, will correspond to $7$-$7$ strings. 
One can see that these $7$-$7$ strings fall into two distinct sectors: large macroscopic spinning strings corresponding to mesons with large spin; 
and small strings captured by the flavor $D7$ fluctuations corresponding to spin 0,1 mesons. These mesons were first studied in \cite{Kruczenski:2003be}.

The mass of the low spin mesons with arbitrary quantum numbers $M$ is of order $m_{M}=m_q/\sqrt{\lambda}$, as opposed to the mass of the high spin mesons 
which is at least of order $m_{M}\lambda^{\frac{1}{4}}$. Therefore in the strong coupling regime we see that higher spin mesons are much more massive than low spin mesons. 
This hierarchy allows us to concentrate on DBI mesons whilst forgetting about the more stringy large spin states. 
Therefore for the mesons of interest, the spectrum can be computed by considering fluctuations, up to quadratic order, of the DBI+CS action of the probe flavor branes. 
Let us consider our $D7$ to be localised at $\vec{z}=(L,0)$ where the scalar fluctuations will be $\vec{z}=(L+\Phi_1,\Phi_2)$. 
In order to have canonical mass dimensions we must re-scale the field to $\Phi_i=2\pi l_s^2\chi_i$. 
Additionally, we have to take into account the fluctuations of the gauge field on the $D7$. 
After considering the quadratic expansion for the flavor branes action, one can see that the scalar wavefunction corresponding to a field of mass $m_{n,l}$ is given by (see \cite{Kruczenski:2003be} for example)

\begin{equation}
\chi_i=e^{p_fx}\Phi_M(r)\mathcal{Y}^{l}\ ;\quad m_M^2=m_{n,l}^2=\frac{2m_q^2}{\lambda}(n+l+1)(n+l+2)\ ,
\end{equation}
where $\mathcal{Y}^{l}$ is the $S^3$ spherical harmonic which specifies the $SU(2)_R\times SU(2)$ quantum numbers of the meson $(\frac{l}{2},\frac{l}{2})$. The function $\Phi_M$ is a radial function with quantum numbers $M=\{n,l\}$ given by

\begin{equation}
\Phi_M=\Phi_{n,l}=\frac{w^{\frac{l}{2}}}{(1-w)^{\frac{l}{2}}}\, _2F_1(-1-l-n,2+l+n,l+2,w)\ ,
\end{equation}
where we have introduced the coordinate $w$ defined through

\begin{equation}
\frac{r^2}{L^2}=\frac{w}{1-w}\ ;\qquad w\in[0,1]\,.
\end{equation}

From the asymptotic behavior of this mode one can see that it is dual to a scalar operator of conformal dimension $\Delta=l+3$, which schematically reads

\begin{equation}
(\tilde{Q}\Phi^lQ)_{\theta\bar{\theta}}=\tilde{\psi}_{\tilde{Q}}\phi^l \psi_Q+\cdots\ ,
\end{equation}
where $\psi_Q,\,\tilde{\psi}_{\tilde{Q}}$ are the fermions in the $Q,\tilde{Q}$ supermultiplets and $\phi$ is the scalar in $\Phi$.

From the eigenmodes of the vector field on the $D7$-brane we get a tower of massive spin-1 $\rho$ mesons whose wavefunctions are (again see \cite{Kruczenski:2003be} for more detail)

\begin{equation}
\rho_{\mu}=\epsilon_{\mu}e^{px}\Phi^{II}_M(r)\mathcal{Y}^l\ ; \quad m_M^2=m_{n,l}^2=\frac{2m_q^2}{\lambda}(n+l+1)(n+l+2)\ ;
\end{equation}
where the polarization vector satisfies the gauge condition $\epsilon\cdot p=0$. 
The $\mathcal{Y}^l$ is the l-th spherical harmonic specifying the $SU(2)_R\times SU(2)$ $(\frac{l}{2},\frac{l}{2})$ representation, whilst $\Phi^{II}_M$ 
is a radial function with quantum numbers $M=\{n,l\}$ given by 

\begin{equation}
\Phi_M^{II}=\Phi_{n,l}^{II}=\frac{w^{\frac{l}{2}}}{(1-w)^{\frac{l}{2}}}\, _2F_1(2+l+n,-1-l-n,l+2,w)\ .
\end{equation}
This wavefunction corresponds to a spin-1 operator of conformal dimension $\Delta=l+3$ schematically given by \cite{Hong:2003jm}

\begin{equation}
\label{components}
(Q^{\dagger}\Phi^lQ-\tilde{Q}\Phi^l\tilde{Q}^{\dagger})_{\theta\bar{\theta}}=q^{\dagger}\phi^l\partial^{\mu}q-\tilde{q}\phi^l\partial^{\mu}\tilde{q}^{\dagger}+\cdots\ ,
\end{equation}
where $q,\,\tilde{q},\,\phi$ stand for the lowest (scalar) components in the $Q,\,\tilde{Q},\, \Phi$ supermultiplets.

Both the scalar and vector meson modes correspond to normalizable fluctuations. However, we can construct the non-normalizable fluctuations 
starting from the same equation of motion. In particular we will be interested in the vector field non-normalizable mode since, as clear from (\ref{components}), 
the $l=0$ case reduces to the flavor current. This current is a global symmetry, exactly as EM is to QCD. 
Therefore we will refer to the ``photon'' as the non-normalizable mode arising from the vector field on the $D7$-branes. 
The flavor symmetry is $SU(N_f)$, but we will choose some $U(1)$ subgroup as our electromagnetic current.
Therefore we will neglect the non-abelian dynamics on the $D7$-branes. 
From this perspective our theory essentially reduces to that of a single flavor D7 brane.
The explicit form of the non-normalizable mode is given by \cite{Hong:2003jm}

\begin{equation}
A_{\mu}=\chi_{\mu}e^{qx} A(r)\mathcal{Y}^0\ , \qquad \chi\cdot q=0\ ,
\end{equation}
where we keep explicit the (trivial) $S^3$ dependence through $\mathcal{Y}^0$. 
However since this spherical harmonic is a constant we will drop it in our computations. Also note that

\begin{equation}
A=\frac{\pi \alpha(1+\alpha)}{\sin(\pi\alpha)}\, _2F_1(-\alpha,1+\alpha,2,w)\ ,\qquad\alpha=\frac{1}{2}(-1+\sqrt{1-\frac{q^2\lambda}{m_q^2}})\ .
\end{equation}

By expanding the effective DBI+CS action on the $D7$-branes to higher orders it is possible to obtain the interacting terms of the meson effective field theory. 
It is to be expected that each such term in that effective field theory is suppressed by extra powers of $N_c$. 
Therefore we will keep the lowest order terms at which we find the desired interaction vertices as the main contribution to the process in which we are interested. 
In our particular case we want to probe the internal structure of mesons with photons. 
Since our photon actually comes from the non-normalizable mode of the vector field on the brane, the lowest order interactions will come from terms in the 
expansion of the DBI+CS which involve two (not necessarily identical) mesons plus a gauge field, which we will interpret as the EM current. 
Clearly at least one of the mesons should be a vector meson in order to contract the indices of the EM current, so from this point of view, it is clear that we 
will find interaction vertices allowing us to compute scalar-vector transition form factors. We will confirm this by direct computation.

\vspace{0.5cm}
\textit{DBI action:}
\vspace{0.2cm}

Starting with the DBI lagrangian for the $D7$-brane, it is convenient to parametrize the fluctuations in terms of the matrix $\epsilon$ in such a way that the DBI reads

\begin{equation}
S=-T_7\int r^3\sqrt{\hat{g}}\sqrt{\rm{det}(1+\epsilon)}\ ; \quad \epsilon^I_J=g^{IL}(h^{\frac{1}{2}}\partial_L\vec{\Phi}\partial_J\vec{\Phi}+2\pi\alpha'F_{LJ})\ ,
\end{equation}
Here capital latin indices run over the worldvolume coordinates of the $D7$, and $\sqrt{\hat{g}}$ is the determinant of the internal unit $S^3$. 
Note that any overall factors of the warp factor cancel out because of having $D7$-branes. 
It is important to stress that the fluctuation metric $g$ depends on the warp factor $h$, which explicitly depends on the fluctuations $\vec{\Phi}$. 
Therefore even though we will expand in powers of $\epsilon$, at each order a further expansion of $g$ is implicit.

To lowest order we find that

\begin{equation}
\label{epsilonexpansion}
\sqrt{\rm{det}(1+\epsilon)} = 1 + \frac{1}{2}Tr(\epsilon) - \frac{1}{4} Tr(\epsilon^2)+\frac{1}{8}(Tr(\epsilon))^2 + \ldots
\end{equation}
Clearly, the linear term will not contribute. From the quadratic terms, to lowest order in the implicit expansion of $g$, we will obtain the quadratic action 
leading to the above wavefunctions. However we will also get extra terms, which in particular contain the interaction lagrangian

\begin{eqnarray}
S^i_{DBI}&=&-T_7(2\pi\alpha')^2\int \sqrt{\hat{g}}r^3\Big\{\frac{LR^4}{(r^2+L^2)^3}\Phi_1F_{\mu\nu}F_{\alpha\beta}\eta^{\mu\alpha}\eta^{\nu\beta}+\frac{2L(r^2+L^2)}{R^4}\Phi_1F_{ri}F_{rj}\hat{g}^{ij}\nonumber\\ &&+\frac{L(r^2+L^2)}{R^4}\Phi_1F_{ij}F_{kl}\hat{g}^{ik}\hat{g}^{jl}\Big\}\ .
\end{eqnarray}
Here latin indices run over the $S^3$, while greek ones are along Minkowski directions.

One can convince oneself that higher orders in the expansion of (\ref{epsilonexpansion}) will contribute to higher point functions, so the expansion in (\ref{epsilonexpansion}) is indeed enough for our purposes.

In the interacting lagrangian we will assume that one of the field strengths corresponds to a non-normalizable gauge field. 
By inspecting the non-normalizable mode above, it is clear that the only contribution will come from the first term -  which in turn requires the other field strength
to be that of the massive vector field. \footnote{The vector field on the $D7$ actually come in 3 modes, out of which we concentrated on the only one which has spin 1. The other modes correspond to scalar fields, and one can convince oneself that they do not couple to the interacting lagrangian above.} Therefore, the vertex on which we will focus is 

\begin{equation}
S^i_{DBI}=-T_7(2\pi\alpha')^3\int \sqrt{\hat{g}}r^3\Big\{\frac{LR^4}{(r^2+L^2)^3}\chi_1F_{\mu\nu}F_{\alpha\beta}\eta^{\mu\alpha}\eta^{\nu\beta}\Big\}\ ,
\end{equation}
where we have extracted the $2\pi\alpha'$ factor in $\Phi_1$ to write the lagrangian explicitly as a coupling to $\chi_1$.

As advertised, our interaction involves a photon, a vector meson and a scalar meson. This structure is deeply connected with the fact the flavors 
(and therefore the mesons) come from fluctuations of a $D7$-brane. In the brane theory the scalars, being real, will not exhibit minimal 
coupling to the vector field. Instead the elements we have to play with are field strengths and derivatives of scalars. 
We may have wondered if we could get a form factor with the same meson for in and out states. However it is clear that we cannot achieve this, 
since that particular interaction could only come from $F_{\mu\nu}$ times some tensor made out of the derivatives of the field, and this vanishes by antisymmetry. 
It is worth noting that if we include a background antisymmetric field, such as a magnetic $B$ field or a worldvolume instanton 
(going to the Higgs phase of the theory \cite {Erdmenger:2005bj, Arean:2007nh}), this restriction can be avoided. It would certainly be interesting
to compare this with our results, which are essentially probing the Coulomb branch of the theory.

\vspace{0.5cm}
\textit{CS action:}
\vspace{0.2cm}

The $AdS_5\times S^5$ background has a non-zero 4-form potential whose electric component is given by

\begin{equation}
C^{(4)}=\frac{\rho^4}{R^4}dx_0\wedge dx^1\wedge dx^2\wedge dx^3\ .
\end{equation}
where $\rho^2=r^2+\vec{z}^2$. 

The relevant coupling in the CS of the flavor $D7$ is now

\begin{equation}
\label{CS}
\frac{2\pi T_7\alpha'}{2}\int C^{(4)}\wedge F\wedge F\ .
\end{equation}
which we can integrate by parts to write as a function of the $G^{(5)}$

\begin{equation}
\frac{2\pi T_7\alpha'}{2}\int A \wedge G^{(5)}\wedge F\ .
\end{equation}
where we denote by $G^{(5)}$ the 5-form field strength derived from $C^{(4)}$, and $A$ is the worldvolume vector field whose corresponding field strength is $F$. 
After some algebra one can see that

\begin{equation}
\label{F5mag}
G^{(5)}=\frac{4R^4}{3(r^2+(L+\Phi_1)^2+\Phi_2^2)^3}\left(\frac{r^4}{4}\omega_3\wedge dx_8\wedge dx_9+r^3x_8 dr\wedge 
\omega_3\wedge dx_9-r^3x_9 dr\wedge \omega_3\wedge dx_8 \right) 
\end{equation}
Since (\ref{CS}) contains two factors of the gauge field, the lowest order which contributes to our interaction vertex will come from a term in $G^{(5)}$ 
containing just one factor of the scalar fluctuation. Clearly this can only arise from the last two terms, where the pull-back of $G^{(5)}$ forces us to 
select the fluctuation through its derivative along the Minkowski directions. It is not hard to convince oneself that finally the relevant CS contribution is

\begin{equation}
S_{CS}^i=\frac{2T_7(2\pi\alpha')^2R^4L}{3}\int  \frac{r^3\sqrt{\hat{g}}}{(r^2+L^2)^3}A\wedge d\Phi_2\wedge F\ .
\end{equation}
Recalling that $d\Phi_2$ actually stands for the derivative along the Minkowski direction, and extracting the $2\pi\alpha'$ dependence from $\Phi_2$, 
we can integrate this by parts to get

\begin{equation}
S_{CS}^i=\frac{T_7(2\pi\alpha')^3R^4L}{3}\int  \frac{r^3\sqrt{\hat{g}}}{(r^2+L^2)^3}\Big\{\chi_2F_{\alpha\beta}F_{\mu\nu}\epsilon^{\alpha\beta\mu\nu}\Big\}\ .
\end{equation}
The lagrangian above demands us to interpret $\chi_2$ as a pseudoscalar, since otherwise the effective meson theory would violate parity. 
In the UV theory we can set the $\theta$ angle to zero, which is dual to taking the RR scalar $C^{(0)}$ to zero. \footnote{Even if we took a pure gauge but 
non-vanishing $C^{(0)}$, it would not couple to this order; suggesting that indeed our effective lagrangian is insensitive to the parity-violating sector of the theory.}
Since in the $N_f/N_c\sim 0$ limit all non-perturbative corrections are switched off, we would not expect any source of parity violation; which demands us to 
consider $\chi_2$ as a pseudoscalar. 

We can provide an additional motivation for this assignment by assuming that the $D7$-branes are localised at a generic point $\vec{L}=(L_1,L_2)$. 
It is then straightforward to repeat the computation above and show that the full interacting lagrangian (DBI+CS) actually reads

\begin{equation}
S^i=T_7(2\pi\alpha')^3R^4\int \frac{r^3\sqrt{\hat{g}} }{(r^2+\vec{L}^2)^3}\Big\{ \vec{L}\cdot\vec{\chi}_V\, F_{\mu\nu}F_{\alpha\beta}\eta^{\alpha\mu}\eta^{\beta\nu}+\frac{1}{3}\vec{L}\cdot\vec{\chi}_A\, F_{\alpha\beta}F_{\mu\nu}\epsilon^{\alpha\beta\mu\nu}\Big\}\,;
\end{equation}
where $\vec{\chi}_V=(\chi_1,\chi_2)$ and $\vec{\chi}_A=\epsilon_{ij}\chi^i=(\chi_2,-\chi_1)$. Thus we see that the sign of the CS term actually depends on the 
choice of skewness of the $(8,9)$ directions. We can consider the interacting lagrangian above just at the level of pure field theory, and suppose now that under
a parity transformation $\vec{x}\rightarrow -\vec{x}$ we should also consider the combined transformation

\begin{eqnarray}
\label{godgivenrule}
(L_1,L_2)& \rightarrow& (L_2,L_1) \nonumber \\ 
(\chi_1,\chi_2)&\rightarrow& (\chi_2,\chi_1)
\end{eqnarray}
Under this transformation it is clear that $\vec{L}\cdot \vec{\chi}_V$ behaves as a scalar, whilst $\vec{L}\cdot\vec{\chi}_A$ picks an extra minus sign 
compensating the minus sign picked up by $F\tilde{F}$. Therefore the transformation (\ref{godgivenrule}) allows for conserved parity. 

We would like to heuristically motivate it yet another way. We could start with the SUGRA background and impose $\vec{x}\rightarrow-\vec{x}$ as a symmetry. 
Clearly the metric is left invariant, however the electric part of the 5-form field strength picks up a minus sign. 
Since $G^{(5)}$ must be self-dual we have to ensure that its magnetic part also picks a minus sign. 
By inspection of (\ref{F5mag}) one can achieve this by reversing the skewness of the $(8,9)$ plane. 
When looking at the linearization of the 5-form one can check that it indeed reduces to (\ref{godgivenrule}). 
If we now choose the particular vacuum $L_1=0, L_2=L$ we see that effectively it is like considering $\chi_1$ as a scalar and $\chi_2$ as a pseudoscalar.
 \vspace{0.7cm}
\\
\\
 The final interaction lagrangian which we will be using reduces to
 
 \begin{equation}
\label{Lint}
S^i=-T_7(2\pi\alpha')^3LR^4\int \frac{r^3\sqrt{\hat{g}}}{(r^2+L^2)^3}\Big\{\chi_1(F^A)_{\mu\nu} (F^{\rho})_{\alpha\beta} \eta^{\mu\alpha}\eta^{\nu\beta}+\frac{1}{3}\chi_2(F^A)_{\alpha\beta} (F^{\rho})_{\mu\nu}\epsilon^{\alpha\beta\mu\nu}\Big\}\,,
\end{equation}
where we have added the superscripts $A$ and $\rho$ to remind the reader that one of the field strengths corresponds to the non-normalizable field 
dual to the photon, while the other corresponds to the in/out $\rho$ meson state.

It is important to note the difference in the measure with respect to the form factors computed in \cite{Hong:2003jm}. 
The additional suppression by $(r^2+L^2)^{-1}$ will be crucial in order to get the expected large $q^2$ behavior of our form factors. 
One can heuristically understand this dependence in much the same spirit as how we motivated the scalar-vector-photon vertex. 
Because of the DBI+CS structure, as discussed, this is the lowest order term we could have. Additionally since the scalar appears without derivatives, 
it could only come from the expansion of a term schematically of the form $hF^2$. In order to get a single power of the scalar 
it should appear in the combination $L\Phi$ in order to have dimensions of $(\rm{length})^2$, however on dimensional grounds, each time this combination appears
it must be suppressed by an extra power of the other dimensionful quantity in the theory, namely the combination $r^2+L^2$. 
Therefore an extra suppression on $(r^2+L^2)^{-1}$ with respect to \cite{Hong:2003jm} is to be expected.

\section{EM transition form factors at strong coupling}

Armed with the interacting lagrangian (\ref{Lint}) we can now turn to the actual problem of computing the electromagnetic form factors. As we discussed, we have to understand $\chi_1$ as a scalar and $\chi_2$ as a pseudoscalar. Even though this pseudoscalar is not the pseudo-Goldstone boson of any broken chiral symmetry, we will call it $\pi^0$, since at least its effective couplings are identical to those of the neutral pion. \footnote{Maybe it would be more convenient to call it $\eta'$, since our theory does not have a chiral non-abelian symmetry but has an approximate UV chiral $U(1)_R$. It is also possible to argue that there should be a pseudoscalar coupling to EM as $\chi F\tilde{F}$.} In turn, the scalar behaves as a scalar neutral meson such as the $f_0$ (or $\sigma$). 

Generically, the situation we will consider is that in figure (1), where the momenta are chosen so that $p+q=p'$.

\begin{figure}[!h]
\centering
\includegraphics[scale=0.5]{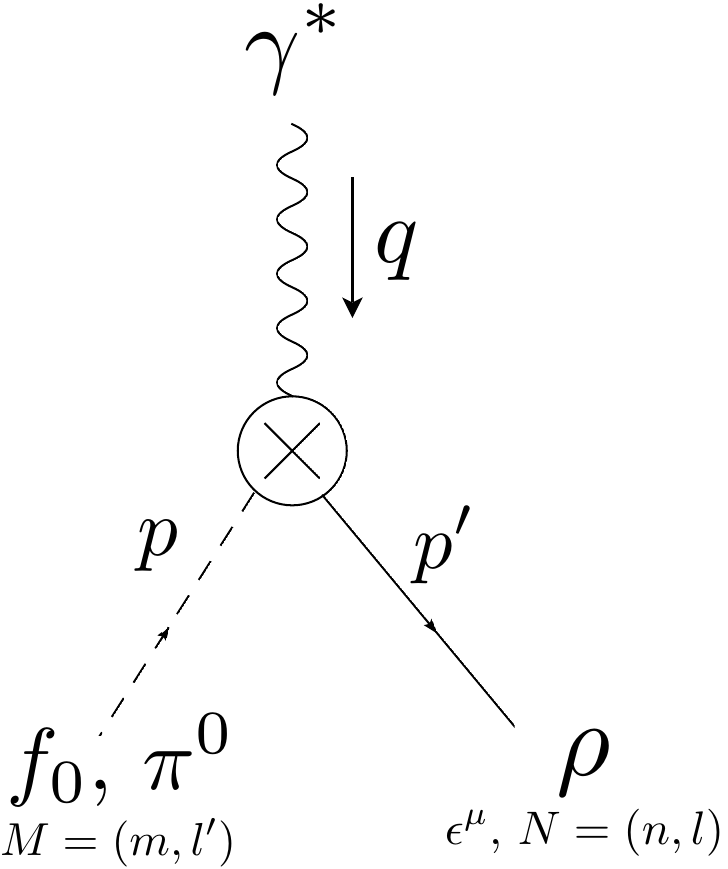}
\caption{Transition form factors for $\pi/f_0 - \rho$ mesons}
\end{figure}

\subsection{$f_0-\rho$ transition form factor}

Using the expressions for the non-normalizable vector field, scalar and vector normalizable modes, it is straightforward to see that

\begin{equation}
\langle f_0,M|J^{\mu}|\rho,\epsilon,N\rangle= 2T_7(2\pi\alpha')^3LR^4\Big(\int \frac{r^3}{(r^2+L^2)^3} A \Phi_M  \Phi_N^{II}\delta_{l,l'}\Big) \Big[(p'\cdot q)\epsilon^{\mu}-(q\cdot\epsilon)p'^{\mu}\Big]\ .
\end{equation}
Here we have already performed the integral over the $S^3$, which is simply

\begin{equation}
\int \sqrt{\hat{g}}\,\mathcal{Y}^l\mathcal{Y}^{l'}=\delta_{l,l'}\ ,
\end{equation}
since the spherical harmonics are orthonormal eigenfunctions of the laplacian on $S^3$.

Let us define the radial integral

\begin{equation}
\label{IM}
I_{n,m,l}(q^2)=\int \frac{r^3}{(r^2+L^2)^3} A\, \Phi_M \Phi_N^{II}\delta_{l,l'}=\int \frac{r^3}{(r^2+L^2)^3} A\, \Phi_{m,l}\Phi_{n,l}^{II}\ .
\end{equation}
It is important to notice that it will be a function of the momentum $q$ of the off-shell photon. Then

\begin{eqnarray}
\langle f_0,M|J^{\mu}|\rho,\epsilon,N\rangle&=& F^{\sigma\rho}_{n,m,l}\,\Big[(p'\cdot q)\epsilon^{\mu}-(q\cdot\epsilon)p'^{\mu}\Big]\ ;\\
F^{\sigma\rho}_{n,m,l}&=&2T_7(2\pi\alpha')^3LR^4I_{n,m,l}(q^2) \ .
\end{eqnarray}
Because of current conservation, $q_{\mu}\langle f_0,M|J^{\mu}|\rho,\epsilon,N\rangle=0$. Let us concentrate on the tensor structure of the form factor. If we go to the rest frame of the vector meson the form factor reduces to

\begin{equation}
\langle f_0|J^{\mu}|\rho\rangle\sim m_{\rho}\Big[-q^0\epsilon^{\mu}-(q\cdot\epsilon)\delta_{\mu}^0\Big]=m_{\rho}\Big[-q^0\epsilon^0\delta^{\mu}_0-q^0\epsilon^i\delta_i^{\mu}+q^0\epsilon^0\delta_0^{\mu}\Big]=-m_{\rho}q^0\epsilon^i\delta_i^{\mu} .
\end{equation}
We might now choose to align $\vec{q}$ with the $z$ direction. Current conservation requires then that $q_{\mu} \langle f_0|J^{\mu}|\rho\rangle=0$, which in turn implies that $\vec{\epsilon}=(\epsilon_x,\epsilon_y,0)$; explicitly showing that the polarization of the vector meson is transverse. 
Therefore, only the vector part of $\epsilon$ gives a non-zero contribution to the transverse part of the off-shell current, reflecting the fact that the 
transition is between a spin 0 and a spin 1 state, and thus should involve the spin 1 part of the current.

For later purposes, let us note that, in the Breit frame where $\vec{p}+\vec{p}'=0$, we will find that all the momenta are of order $q^2$.
\footnote{In that frame $\vec{p}'=-\vec{p
}=\frac{\vec{q}}{2}$, and we may choose $\vec{q}=(0,0,q)$. A boost along the $z$ direction connects this frame with the rest frame of the vector meson. 
Since in the rest frame $\epsilon=(0,\vec{\epsilon}_{\bot},0)$, the vector meson will be also transverse in the Breit frame. If we have ultra-relativistic hadrons, $q_0\sim 0$, then $p=(\frac{|\vec{q}|}{2},-\frac{\vec{q}}{2})$, $p'=(\frac{|\vec{q}|}{2},\frac{\vec{q}}{2})$, $q=(0,\vec{q})$.} In that case, we see that roughly speaking, the large $q^2$ behaviour of the matrix element will be given by $I_{n,m,l}(q^2) q^2$.

\subsection{$\pi-\rho$ transition form factor}
 
It is also straightforward to evaluate the CS interaction term. It reads

\begin{eqnarray}
\langle \pi^0,M|J^{\mu}|\rho,\epsilon,N\rangle&=& F^{\pi\rho}_{n,m,l}\, \epsilon^{\mu\nu\alpha\beta}\epsilon_{\nu}q_{\alpha}p'_{\beta}\ ; \\ F^{\pi\rho}_{n,m,l}&=&\frac{4}{3}T_7(2\pi\alpha')^3LR^4I_{n,m,l}(q^2)\ ;
\end{eqnarray}
where $I_{n,m}(q^2)$ is the same integral (\ref{IM}) as above. One can check that also current conservation is satisfied and $q_{\mu}\langle \pi^0,M|J^{\mu}|\rho,\epsilon,N\rangle=0$.

Again we can go to the rest frame of the vector meson where

\begin{equation}\label{pirho}
\langle \pi^0|J^{\mu}|\rho\rangle\sim m_{\rho}\epsilon^{\mu\nu\alpha0}\epsilon_{\nu}q_{\alpha}\sim m_{\rho}\epsilon^{\mu i j0}\epsilon_iq_j\\ 
\end{equation}
and we again see that only the vector part of the polarization of the vector meson is involved - and therefore only the spin 1 part of the current is involved in the
interaction.

Note that again, in the Breit frame, the magnitude of the whole matrix element for large $q^2$ will be of the order $I_{n,m,l}(q^2)q^2$.
We also point to the results obtained in \cite{  Brodsky:2003px, deTeramond:2005su, Brodsky:2006uqa, Brodsky:2007hb, Brodsky:2008pg, Brodsky:2008pg} for the form factor, albeit in a slightly different set-up.

\subsection{$q/\vec{x}$ dependence of the form factors}

By inspection of the two form factors, we have that $F^{\pi\rho}_{n,m,l}=\frac{2}{3}F^{\sigma\rho}_{n,m,l}$.  Since in addition $F^{\sigma\rho}_{n,m,l}\sim\, I_{n,m,l}$, we will loosely identify $I_{n,m,l}$ with the form factors of interest. The fact that both form factors are proportional one to the other should be due to SUSY, since both the two scalars are in the same $\mathcal{N}=2$ massive supermultiplet.

In order to go further, we should study $I_{n,m,l}$. Following the method suggested in \cite{Hong:2003jm}, we can do the $j$-th integration of $A$ with respect to $w$ - which we will call $a_j$. 
Recall that $A$ is a function of the photon momentum, so for integration we must recall that $a_j=a_j(w,q^2)$. 
Denoting by $F$ the rest of the hypergeometric functions under the integral in (\ref{IM}), we can iteratively integrate this by parts to obtain

\begin{eqnarray}
I_{n,m,l}=\int A\, F=\int \partial_{\omega}a_1F&=&(a_1 F)|^{w=1}_{w=0}-\int a_1\partial_{\omega}F=\nonumber \\ &=&(a_1 F)|^{w=1}_{w=0}-(a_2\partial_{\omega} F)|^{w=1}_{w=0}+\int a_2\partial_{\omega}^2F\ =\cdots\ .
\end{eqnarray}
It can be easily checked that $a_j|_{w=0}=0$. The crucial observation then is that only a finite number of derivatives of $F$ are 
non-vanishing when evaluated at $w=1$. 
We find that the last non-zero derivative is the $j_{max}=2l+n+m+3$. 
Analogously one can check that the first non-zero derivative is the $j_{min}=l+2$. This way we see that indeed 

\begin{equation}
\label{expansion}
I_{n,m,l}=\sum_{j_{min}}^{j_{max}}\,(-1)^j\, a_{j+1}(q^2)\,\partial^{j}_wF\ ;
\end{equation}
where both $a_{j+1}$ and $\partial^j_wF$ are evaluated at $w=1$. By iteration one can, in principle, determine the value of the integral by obtaining all the
higher order coefficients.

We could take a seemingly different approach and make direct use of VMD (see appendix A). It is possible then to rewrite (\ref{IM}) as

\begin{equation}
\label{IVMD}
I_{n,m,l}=\frac{m_q^2}{\lambda}\sum_{k=0}^{k_{max}} \frac{f_{k,0} \,\mathcal{R}_{n,l,m,l,k,0}}{q^2+m_{k,0}^2}\ ;\quad \mathcal{R}_{n,l,m,l,k,0}=\int\frac{r^3}{(r^2+L^2)^3}\Phi_{m,l}\Phi^{II}_{n,l}\Phi^{II}_{k,0}\ ,
\end{equation}
where $\mathcal{R}_{n,l,m,l,k,0}$ is proportional to the coupling constant between hadrons with the specified quantum numbers, and $f_{k,0}$ is the decay constant of the vector meson with quantum numbers $(n,l=0)$. 
A priori $k$ in (\ref{IVMD}) should take values in the range $k\in[0,\infty)$. 
However it can be checked that $\mathcal{R}_{n,l,m,l,k,0}=0$ for $k> k_{max}$, with $k_{max}=2l+n+m+2$, which truncates the sum and ensures its finiteness. 
Additionally we should point out that the minimal $k$ depends on the particular choice of $(n,m)$. For example, for $n=m$ the sum extends all the way down to $k=0$.

Expression (\ref{IVMD}) should coincide with (\ref{expansion}). However, it has a compelling interpretation since it explicitly 
allows us to see the dependence on the decay and coupling constants. Indeed expression (\ref{IVMD}) is the manifestation of VMD in gauge/gravity duality. 
It allows us to regard the photon-hadron interaction as a photon going into a vector meson which is indeed the one which interacts with the hadrons, 
in accordance with the vector meson dominance principle. The pictorial representation is in figure (2).

\begin{figure}[!h]
\centering
\includegraphics[scale=0.5]{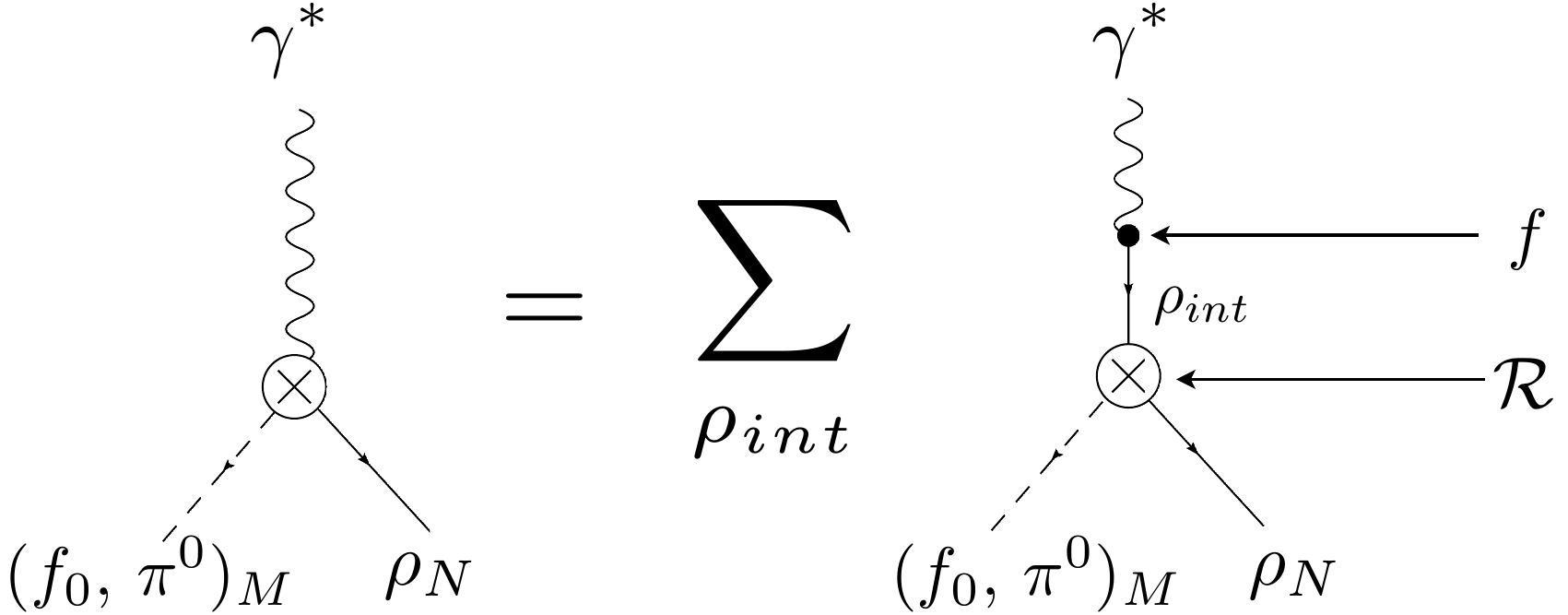}
\caption{Vector meson dominance.}
\end{figure}
It should be pointed that even though VMD naturally falls out in the gravity construction, universality does not generically hold. 
See \cite{Hong:2004sa} for further details on this.

The necessary agreement between (\ref{IVMD}) and (\ref{expansion}) requires some, a priori, non-obvious relation between masses, decay constants and 
coupling constants. We will make explicit use of some of these properties below.

It is expected that form factors in position space carry information about the charge distribution of the hadrons. 
This interpretation is most straightforward for diagonal form factors (\textit{i.e.} form factors in which the \textit{in} meson is 
identical to the \textit{out} meson). However in the case of off-diagonal form factors (transition form factors) 
we could think of it as the distribution of charge at the interaction point.

There are, however, subtleties in how one should extract this kind of information in position space from a form factor computed in momentum space. 
For example, in a non-relativistic system the 3-dimensional Fourier transform of the form factor with respect to $\vec{q}$ 
would give the spatial charge distribution. However since our mesons have a very high binding energy, one should expect our system to be highly relativistic. 
It has been argued in \cite{Hong:2003jm} that, in order to have the right probabilistic interpretation (as well as a connection to generalized PDF, 
see for example \cite{Radyushkin:1997ki, Burkardt:2002hr, Ralston:2001xs}), it is natural to switch to the large momentum frame, and interpret the photon as probing the transverse 
structure of the hadron. 
As suggested in \cite{Hong:2003jm}, we might then consider aligning the initial hadron momentum along the $z$ direction, and boosting the system to 
large momentum along it. Then choosing $q=(0,\vec{q}_{\bot},0)$ we can perform a 2-dimensional Fourier transform

\begin{equation}
FT_2\Big(f(q)\Big)=\frac{1}{2\pi}\int d^2\vec{q}_{\bot}e^{i\vec{q}_{\bot}\vec{x}_{\bot}}f(q)\, ;\quad FT_2\Big(\frac{1}{q^2+m_{k,0}^2}\Big)=K_0(m_{k,0} r)\ ,
\end{equation}
where $K_0(x)$ is the corresponding Bessel function. The function obtained by means of this 2d Fourier transform should be interpreted as a charge 
density in the transverse space parametrized by the transverse radius $r$. Restricting to the case at hand we have, from (\ref{IVMD})

\begin{equation}
\label{Iposition}
I_{n,m,l}=\frac{m_q^2}{\lambda}\sum_k^{k_{max}} f_{k,0}\, \mathcal{R}_{n,l,m,l,k,0}\, K_0(m_{k,0} r)\ .
\end{equation}
In the case of diagonal form factors, one possible definition of the size of the hadron is

\begin{equation}
\label{raverage}
\langle r^2\rangle=4\frac{\partial}{\partial q^2} F_{diag}(q^2) \rvert_{q^2=0}\ ,
\end{equation}
where the factor of 4 accounts for the fact that this is a transverse (2-dimensional) charge distribution. 
We will use this definition and interpret it as a measure of the size of the region where the interaction takes place. Then using (\ref{IVMD}) we have, for the $\sigma\rho$ transition form factor

\begin{equation}
\langle r^2_{\sigma\rho}\rangle=8T_7(2\pi\alpha')^3LR^4 m_q^2\lambda^{-1}\sum_k^{k_{max}}\frac{f_{k,0}\, \mathcal{R}_{n,l,m,k,0}}{m_{k,0}^4}\ .
\end{equation}
Unfortunately we have been unable to explicitly perform the above sum. 

Finally let us note that, from the relation between $F^{\sigma\rho}$ and $F^{\pi\rho}$, we have $\langle r^2_{\sigma\rho}\rangle=\frac{3}{2}\langle r^2_{\pi\rho}\rangle$,
which follows trivially from the definitions of the two form factors.

\vspace{0.5cm}
\textit{Large $q^2$ behavior:}
\vspace{0.2cm}

Let us concentrate on the large $q^2$ behavior of (\ref{IM}). The $j$-th integration of $A$, $a_j$, behaves like

\begin{equation}
a_j\rightarrow4^jj!\Big(\frac{m_q^2}{\lambda q^2}\Big)^{j+1}\ .
\end{equation}
It is clear that the large $q^2$ behavior of (\ref{IM}) will be controlled by the first non-zero derivative of $F$ at $w=1$ (at $w=0$ $a_j$ vanishes) 
since it will be the least suppressed term. As anticipated, the first non-zero derivative is the $l+2$. 
Therefore we see that at large $q^2$, the integral (\ref{IM}) runs like

\begin{equation}
I_{n,m}\rightarrow C_{n,m,l} \Big(\frac{m_q^2}{\lambda q^2}\Big)^{3+l}=C_{n,m,l} \Big(\frac{m_q^2}{\lambda q^2}\Big)^{\Delta}\ ,
\end{equation}
where $C_{n,m,l}$ is a numerical coefficient depending on $n,m,l$. Note that the $q^2$ dependence is completely independent of $n,m$, 
and just relies on the conformal dimension of the operators involved. 
This is connected to the fact that the theory flows to a UV conformal point and so we see that conformal invariance alone governs the structure of the form factors.

We could just as well use the alternate expression (\ref{IVMD}). Expanding (\ref{IVMD}) for large $q^2$ we have

\begin{equation}
I_{n,m,l}\rightarrow I_{n,m,l}=\frac{m_q^2}{\lambda}\sum_j \frac{(-1)^{j}}{(q^2)^{j+1}}\Big(\sum_k f_{k,0}\, \mathcal{R}_{n,l,m,l,k,0}\, (m_{k,0})^{2j}\Big)\ .
\end{equation}
We can re-write this as

\begin{equation}
I_{n,m,l}\rightarrow I_{n,m,l}=\sum_j (-1)^{j}4^jj!\Big(\frac{m_q^2}{\lambda\, q^2}\Big)^{j+1}\Big(\frac{1}{4^j j!}\,\sum_k f_{k,0}\, \mathcal{R}_{n,l,m,l,k,0}\, (\frac{\sqrt{\lambda}\,m_{k,0}}{m_q})^{2j}\Big)\ ,
\end{equation}
so we conclude that the coefficients in (\ref{expansion}) can be written as

\begin{equation}
\partial_w^jF|_{w=1}=\frac{1}{4^j j!}\,\sum_k^{k_{max}} f_{k,0}\, \mathcal{R}_{n,l,m,l,k,0}\, \left(\frac{\sqrt{\lambda}\,m_{k,0}}{m_q}\right)^{2j}\ .
\end{equation}

This allows us to write the leading term for large $q^2$ fixing the $C_{n,m,l}$ above

\begin{equation}
I_{n,m,l}\rightarrow (-1)^{l+2}\Big(\sum_k^{k_{max}} \,\Big(\frac{\lambda\,m_{k,0}^2}{m_q^2}\Big)^{(l+2)}\,f_{k,0}\, \mathcal{R}_{n,l,m,l,k,0}\Big)\Big(\frac{m_q^2}{\lambda q^2}\Big)^{\Delta}\ .
\end{equation}

Since from (\ref{expansion}) we know that the first non-zero term is that with $1/(q^2)^{l+3}$, we conclude that

\begin{equation}
\label{identity}
\sum_k f_{k,0}\, \mathcal{R}_{n,l,m,l,k,0}\, (m_{k,0})^{2j}=0\,\quad \forall j<l+2\ .
\end{equation}
This is an important constraint on the algebraic structure.
For later purposes, let us consider the function ($r\ne 0$)

\begin{equation}
h_j(r)=\sum_k f_{k,0}\, \mathcal{R}_{n,l,m,l,k,0}\, (m_{k,0})^{2j}\,\log(\frac{m_{k,0}r}{2})\ ,
\end{equation}
Taking its $r$ derivative we have

\begin{equation}
\frac{dh_j(r)}{dr} \sim \frac{1}{r}\sum_k f_{k,0}\, \mathcal{R}_{n,l,m,l,k,0}\, (m_{k,0})^{2j}\ ;
\end{equation}
so that by using (\ref{identity}) we see that when $j<l+2$, $h_j$ is actually a constant.

Let us analyze the form factor in position space. As argued above in order to have a sensible physical interpretation we must boost the system 
to the infinite momentum frame and assume $q=(0,\vec{q}_{\bot},0)$. Then we will Fourier transform to obtain a function of the transverse size $r$. 
The large $q^2$ region corresponds to small $r$. Expanding (\ref{Iposition}) and re-writing it in a suitable form, we see that for $r\sim 0$ 

\begin{equation}
I_{n,m,l}\rightarrow \frac{m_q^2}{\sqrt{\lambda}}\sum_k\sum_j \, \frac{f_{k,0}\,\mathcal{R}_{n,l,m,l,k,0}}{2^{2j}j!}\, \big\{\psi(j)-\log\big(\frac{m_{k,0} r}{2}\big)\big\}\big(m_{k,0}r\big)^{2j}\ ,
\end{equation}
where $\psi(j)$ are numerical coefficients depending on $j$. The term dominating the sum above will be the one with the lowest exponent for $r$. Using (\ref{identity}) and the fact that $h_j$ is constant for any $j<l+2$, we conclude that, up to a constant (which on physical grounds must be zero), the small $r$ dependence in transverse space is

\begin{equation}
F^{\sigma\rho}_{n,m,l}\sim F^{\pi\rho}_{n,m,l}\sim I_{n,m,l}\sim r^{2(l+2)}\log r\sim (r^2)^{\Delta-1}\log r\ ,
\end{equation}
where we have re-written the $r$ power in terms of the conformal dimension of the operators involved. 
Note that the scale in position space at large $q^2$ is set by $1/m_{l+2,0}$.

\vspace{0.5cm}
\textit{General behavior going towards the IR:}
\vspace{0.2cm}

Let us now analyse the IR behaviour, \textit{i.e.} the small $q^2$ region. In position space this corresponds to the asymptotically large $r$ region. 
From the asymptotic behavior of the Bessel function we see that

\begin{equation}
I_{n,m,l}\rightarrow\frac{\sqrt{\pi} m_q^2}{\sqrt{2}\lambda}\sum_k f_{k,0}\, \mathcal{R}_{n,l,m,l,k,0}\, \frac{e^{-m_{k,0}r}}{\sqrt{m_{k,0}r}}\sim \frac{\sqrt{\pi} m_q^2}{\sqrt{2}\lambda} f_{\hat{k},0}\, \mathcal{R}_{n,l,m,l,\hat{k},0}\, \frac{e^{-m_{\hat{k},0}r}}{\sqrt{m_{\hat{k},0}r}}\ .
\end{equation}
Where $\hat{k}$ is the lowest $k$ for which $\mathcal{R}_{n,l,m,l,k,0}$ does not vanish. As we pointed out, this minimal $\hat{k}$ depends on the particular choice of $(n,m)$, which in turn sets the scale $1/m_{\hat{k},0}$ of the measured charge distribution in position space for small $q^2$.

\subsection{Field theory expectations for the transition form factors}

As we have discussed, the UV of our theory is described by a conformal point. Therefore we expect the large $q^2$ behavior of our form factors 
to be controlled purely by conformal invariance, in much the same way as in QCD - where asymptotic freedom is responsible for the 
vanishing beta function at large $q^2$. However in that case the theory is weakly coupled 
and one can make use of perturbative tools to study the behavior of diverse processes at large $q^2$ \cite{Brodsky:1974vy, Matveev:1972gb}
(see \cite{Chernyak:1983ej} for an exhaustive review).

Rather than looking directly to form factors, it is useful to consider the full matrix element, $i.e.$ taking into account the scaling of the tensor structure.\footnote{For example, our matrix elements are schematically $I(q^2)\,\epsilon^{\mu\nu\alpha\beta}q_{\alpha}q_{\beta}\sim I(q^2)\,q^2$.} 
In the Breit frame, where all the momenta are of order $q$, we can identify the $q$-dependence of our matrix element (recall equation (\ref{pirho})) as

\begin{equation}
\langle \pi^0, f_0|J^{\mu}|\rho\rangle \sim \frac{1}{(q^2)^{\Delta-1}}\ .
\end{equation}
On the other hand, for a conformal field theory at weak coupling, the expected scaling for the transition form factor between a hadron $h_1$ of helicity $s_1$ and a hadron $h_2$ of helicity $s_2$ is  \cite{Chernyak:1983ej}

\begin{equation}
\label{scaling1}
\langle h_1,s_1|J^{\mu}|h_2,s_2\rangle\sim \frac{1}{q^{2n-3+|s_1-s_2|}}\ ;
\end{equation}
where $n$ is the number of partons. Additionally, (\ref{scaling1}) requires us to impose the selection rule that current helicity is given by $\lambda=s_1+s_2$. 
We can can heuristically understand this formula in a free parton model. Assume that $h_1,\,h_2$ are composed of $n$ partons, each carrying a fraction of the total 
momentum $q$ of the hadron. The off-shell photon would strike one of them which, in the Breit frame, forces the struck parton to recoil. 
Since we are looking into elastic processes, for the hadron not to break we require the struck parton to emit a gluon to force the other partons to recoil.
After power counting in this naive parton model it is easy to see that one recovers (\ref{scaling1}).

From (\ref{scaling1}) it is also clear that form factors in which helicity change is involved are suppressed by additional powers of $q$ \cite {Vainshtein:1977db, Brodsky:1981kj}. 
It is easy to understand this in the naive parton model. The reason is that the vector boson vertex does not change helicity unless the partons are massive. In that case helicity flipping processes are 
suppressed by an extra power of $m/q$.

The discussion above is not limited to weak coupling, since in the end the is tied to conformal invariance ($i.e$. naive dimension counting as if the beta function was zero). Indeed, it can be extended to strong coupling by replacing $n$ by the twist of the lowest twist operator capable of creating \emph{both} hadrons \cite {Polchinski:2001tt, Polchinski:2002jw}. 
In the case at hand we have a spin 0 hadron (conformal dimension $\Delta$) whose twist is $\tau_{S=0}=\Delta$, and a spin 1 hadron (conformal dimension $\Delta$) 
whose twist is $\tau_{S=1}=\Delta-1$. Even though $\tau_{S=1}$ is smaller, the lowest twist operator capable of creating both hadrons has $\tau=\Delta$. 
Thus we can extend (\ref{scaling1}) to strong coupling by replacing $n\rightarrow \Delta$.
 
It remains to discuss the helicities of the in and out hadrons. We saw that in both the $\pi^0-\rho$ and $f_0-\rho$ cases the transition was purely transverse. 
This means that $s_1=0$ for the $f_0 / \pi^0$ while $s_2=\pm 1$ for the $\rho$. According to the selection rule this implies that the part of the current 
involved in the transition is the $\lambda=\pm 1$ part, which is indeed what we found (we will see additional consequences of this when we study the full amplitude). 
Therefore on general grounds, we expect the matrix element to scale like

\begin{equation}
\label{scaling}
\langle h_1,s_1|J^{\mu}|h_2,s_2\rangle\sim \frac{1}{q^{2\Delta-3+1}}=\frac{1}{(q^2)^{\Delta-1}}\ ,
\end{equation}
which is indeed the scaling we obtained.

Note that in order to get this precise scaling the extra suppression by $(r^2+L^2)^{-1}$ in (\ref{Lint}) is crucial. 
At this point it is instructive to compare with the form factors computed in \cite{Hong:2003jm}. For simplicity let us consider the spin 0 case in that paper. \footnote{With 
a little bit of more work one can argue the same is true for the other form factors.}
The corresponding integral leading to the form factor was very similar to (\ref{IM}), but without the extra suppression by $(r^2+L^2)^{-1}$. 
This has the non-trivial effect of making the matrix element scale with an extra power of $1/q^{-1}$ (technically it is due to the fact that the first non-zero 
derivative of the equivalent $F$ would appear one order beyond). On the other hand these form factors are between spin 0 states, and thus we expect that the 
extra suppression due to helicity flip in (\ref{scaling}) to be absent. This justifies the extra power of $1/q^{-1}$. 
We can now re-analyse the appearance of this form factor in view of these results. 
The scalars of the theory are real implying that there will be no minimal coupling to the vector field on the brane. 
Thus the only possible trilinear combination is the one we obtained which, due to dimensional reasons, requires the extra suppression with $(r^2+L^2)^{-1}$. 
Now we re-discover that the dual statement is that the theory recovers conformal invariance in the UV, which dictates the scaling of the form factors.

\subsection{$\gamma^*\gamma^*\rightarrow \pi^0$, $F^{\pi\rho}$ and VMD}

The interacting lagrangian (\ref{Lint}) allows us to study the process $\gamma^*\gamma^*\rightarrow \pi^0$ by considering the two vector fields 
to be non-normalizable modes. To be more precise, we will consider the process shown in figure 3.

\begin{figure}[!h]
\centering
\includegraphics[scale=0.4]{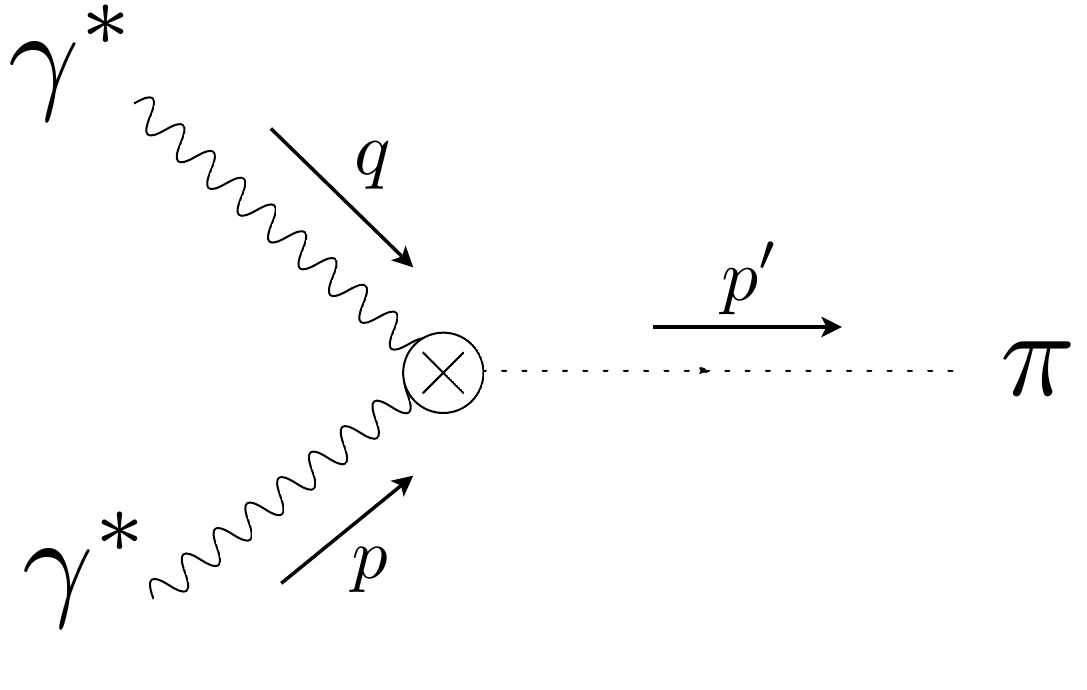}
\caption{$\gamma^*\gamma^*\rightarrow \pi$ process to be considered.}
\end{figure}
By using the CS interaction (\ref{Lint}), it is straightforward to see that this amplitude is given by

\begin{equation}
i\mathcal{M}_{\gamma^*\gamma^*\rightarrow\pi}=\frac{4}{3}T_7(2\pi\alpha')^3LR^4\hat{I}_M(q^2,p^2)\Big(\int \sqrt{\hat{g}}\, Y^l\Big)\big[ \epsilon^{\mu\nu\alpha\beta}\epsilon_{\mu}(q)\epsilon_{\nu}(p)q_{\alpha}p_{\beta}\big]\ ,
\end{equation}
where now $\hat{I}_M(q^2,p^2)$ is given by

\begin{equation}
\label{Ihatgen}
\hat{I}_M(q^2,p^2)=\int\frac{r^3}{(r^2+L^2)^3}A(q^2)A(p^2)\Phi_M\ .
\end{equation}
and $\Phi_M$ is the radial wavefunction of the $\pi^0$ with quantum numbers $M=(n,l)$. 

We can go to the rest frame of the final hadron, where we have $\vec{q}=-\vec{p}$, $q_0+p_0=m_{\pi}$. 
It is straightforward to check that the tensor structure here reduces to $m_{\pi}\epsilon^{ijk0}\epsilon_i(q)\epsilon_j(p)q_k$.

Forgetting for a while about the angular integral, let us define the following form factor

\begin{equation}
F^{\gamma^*\pi}_n=\frac{4}{3}T_7(2\pi\alpha')^3LR^4\hat{I}_{M}(q^2,p^2)\ .
\end{equation}
Using the VMD decomposition (see appendix A) we have

\begin{equation}
\hat{I}_{M}(q^2,p^2)=\frac{m_q^2}{\lambda}\sum_{n'}\frac{f_{n',0}}{p^2+m_{n',0}^2}\int \frac{r^3}{(r^2+L^2)^3}\Phi_{n,l}\Phi^{II}_{n',0}A(q^2)\ ;
\end{equation}
where $f_{n,0}$ is the decay constant of the vector meson with quantum numbers $n,\,l=0$. We can interpret this as in figure 4.

\begin{figure}[!h]
\centering
\includegraphics[scale=0.4]{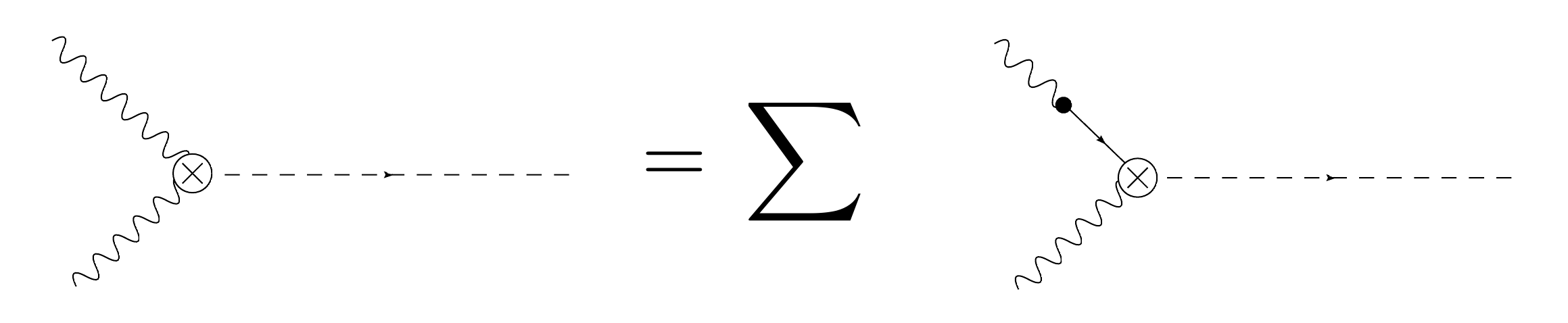}
\caption{$\gamma^*\gamma^*\rightarrow \pi$ after using VMD.}
\end{figure}
Each term in the sum is then the transition form factor for $\rho\pi$, with the caveat that the intermediate vector meson has $l=0$, 
which follows trivially from VMD since the photon is the non-normalizable mode with $l=0$ of the vector field - and as such can only mix with 
vector mesons of $l=0$. This suggests that we interpret the integral over the angular coordinates as

\begin{equation}
\int\sqrt{\hat{g}}\,\mathcal{Y}^l\mathcal{Y}^0\ ;
\end{equation}
where the $\mathcal{Y}^0$ spherical harmonic would correspond to the intermediate vector meson (which has $l=0$). Then the normalisation condition requires

\begin{equation}
\int\sqrt{\hat{g}}\,\mathcal{Y}^l\mathcal{Y}^0=\delta_{l,0}\ ;
\end{equation}
so the final state will only contain the $l=0$ $\pi^0$ meson. 
Therefore we can define the full form factor as $F^{\gamma^*\pi}_n=\frac{4}{3}T_7(2\pi\alpha')^3LR^4\hat{I}_{n,0}(q^2,p^2)$, 
where we make explicit the fact that only the $l=0$ mode contributes. Explicitly

\begin{equation}
\label{Ihat}
\hat{I}_{M}(q^2,p^2)=\frac{m_q^2}{\lambda}\sum_{n'}\frac{f_{n',0}}{p^2+m_{n',0}^2}\int \frac{r^3}{(r^2+L^2)^3}\Phi_{n,0}\Phi^{II}_{n',0}A(q^2)=\frac{m_q^2}{\lambda}\sum_{n'}\frac{f_{n',0}\, I_{n,n',0}(q^2)}{p^2+m_{n',0}^2}\ .
\end{equation}
We can re-write our form factor as

\begin{equation}
\label{disp}
F^{\gamma^*\pi}_n=\frac{f_{0,0}m_q^2\lambda^{-1}F^{\rho\pi}_{0,0,0}}{p^2+m_{0,0}^2}+\int_0^{\infty}ds\, \frac{\rho^h}{s+p^2}\ .
\end{equation}
where we have separated out the contribution of the $\rho$ meson with quantum numbers $(0,0)$, \textit{i.e.} the lowest one in the KK-tower. This is formula is analogous to the one obtained in QCD arising from VMD (see for example \cite{Khodjamirian:1997tk}).

The spectral density reads in this case
\begin{equation}
\rho^h=8cT_7(2\pi l_s^2)^3R^4Lm_q^2\lambda^{-1}\sum_{m'\ne 0} f_{m',0} \delta(s-m_{m',0}^2)\int \frac{r^3}{(r^2+L^2)^3}f_{n,0}\Phi^{II}_{m',0}A(q^2)\ .
\end{equation}
We can use once again the decomposition formula, and write

\begin{equation}
\label{density}
\rho^h=8cT_7(2\pi l_s^2)^3R^4Lm_q^4\lambda^{-2}\sum_{m'\ne 0}\sum_{m''} \frac{f_{m',0}f_{m'',0}}{q^2+m_{m'',0}^2}\mathcal{R}_{n,0,m',0,m'',0} \delta(s-m_{m',0}^2)\ .
\end{equation}

The expression (\ref{density}) for the spectral density follows from vector meson dominance. The interaction $\gamma^*\gamma^*\pi^0$ (or  $\gamma^*\gamma^*\sigma$) can be seen as $\gamma^*\rightarrow \rho$ and $\rho\rho\pi^0$.

\begin{figure}[!h]
\centering
\includegraphics[scale=0.4]{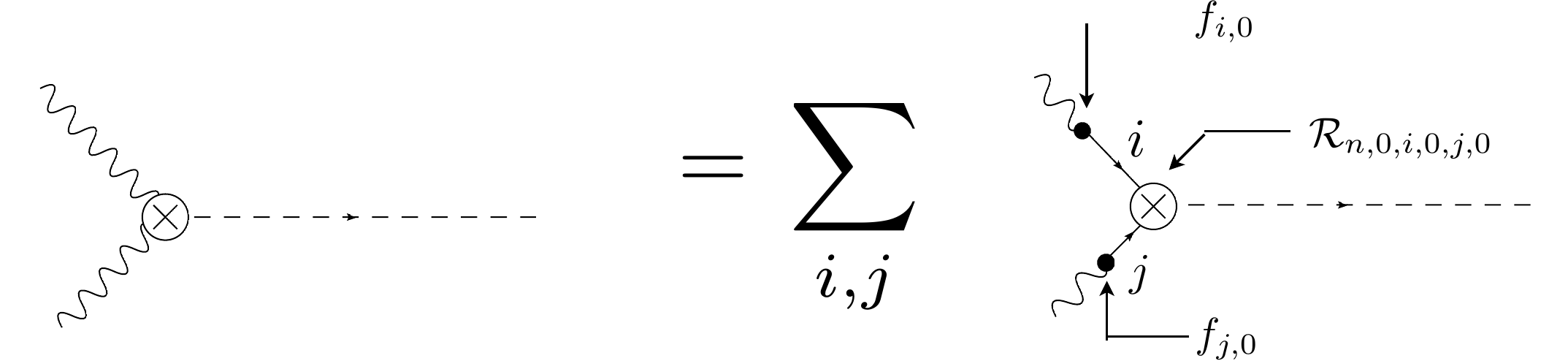}
\caption{$\gamma^*\gamma^*\rightarrow \pi$ in the light of VMD.}
\end{figure}

Then after separating out the lowest mass state, we have that the spectral density is just the sum over higher mass states.

It is interesting to look at the large momentum behavior of the above form factor. Note that, in fact, we could switch the scalar with the pseudoscalar here, and the calculation proceeds in exactly the same fashion albeit with an additional factor
of $2/3$.
From (\ref{Ihat}) we see that for large $q^2$

\begin{equation}
\label{largeq2}
\hat{I}_{n,n',0}\rightarrow m_q^2\lambda^{-1} \sum_{n'}\frac{f_{n',0}}{p^2+m_{n',0}^2}\Big(\sum_k^{k_{max}} \,\Big(\frac{\lambda^2\,m_{k,0}^4}{m_q^4}\Big)\,f_{k,0}\, \mathcal{R}_{n,0,n',0,k,0}\Big)\Big(\frac{m_q^2}{\lambda q^2}\Big)^{3}\\ .
\end{equation}

From (\ref{largeq2}) we can extract the relevant behavior when one of the virtualities is large and the other is small to 
obtain $F^{\gamma^*\pi}\sim 1/(q^2)^3$. We can cross-check this result by using the iterative integration method of section 4. We would like to note that it is straightforward to reapeat a similar computation with the scalar meson getting the same result up to a numerical factor. 

It is instructive to consider the corresponding process with a vector meson as the final state, \textit{i.e.} the process $\gamma^*\gamma^*\rho$. 
This can be obtained from (5.43) in \cite{Hong:2003jm} if we assume the photon is valued in $SU(N)$. 
The crucial difference would be that the analog of (\ref{Ihat}) now reads

\begin{equation}
\tilde{\hat{I}}_{M}(q^2,p^2)=\frac{m_q^2}{\lambda}\sum_{n'}\frac{f_{n',0}}{p^2+m_{n',0}^2}\int \frac{r^3}{(r^2+L^2)^2}\Phi^{II}_{n,0}\Phi^{II}_{n',0}A(q^2)\ .
\end{equation}
Restricting this to the case of one (almost) on-shell photon and the other with large (virtual) momentum ($p^2\sim 0$, $q^2\gg 1$), we can 
transplant the results of \cite{Hong:2003jm}, where it was shown that the integral scales with the second power of $1/q^2$. 
Therefore when both photons and vector meson are polarised in the transverse direction, we find 
that $F^{\gamma^*\rho}\sim 1/(q^2)^2$. 

Let us return to the pseudoscalar (or scalar) case ($\gamma^*\gamma^*\pi^0$ or $\gamma^*\gamma^*\sigma$) assuming one photon 
with large virtuality and the other almost on-shell. 
We found that the form factor scales like $1/q^6$. 
This scaling is different to that obtained in QCD,  where the $\gamma^*\gamma^*\pi$ form factor goes like $1/q^2$. 
In principle one would expect these two results to match due to conformal invariance. Nevertheless, the $\gamma^*\gamma^*\pi$ is more subtle than the 
form factors discussed above. In pQCD (perturbative QCD) it is dominated by the one-quark propagator (see for example \cite{Chernyak:1983ej}), 
which in turn arises from the fact that the $\pi$ meson is a 2-quark bound state. By comparison with the case of the form factors, 
we see that in order to compare weak coupling results with strong coupling results one needs, at least\footnote{The $\gamma^*\gamma^*\pi$ channel is more 
sensitive to details of the theory than the form factors. It might be that our theory, being non-perturbatively trivial, simply has a different structure than QCD.} 
to replace $\tau\leftrightarrow n$. As opposed to the form factors, in the case at hand the fact that the selection rule sets $l=0$ 
obscures the identification of $\tau$. However we can perform the integrals above before taking $l=0$.  
One can see that, considering (\ref{Ihatgen}), in order to have a well-behaved integral we have to restrict ouselves to even values of $l$. 
Under that assumption, one can check that $\hat{I}\sim1/q^{l+6}$, which upon taking $l=0$ coincides with the result obtained using VMD. 
Since $l$ has to be even, we can re-write it as $l=2l'$ in such a way that the integral goes like $1/(q^2)^{l'+3}$. 
Defining a new twist operator $\tau=l'+3$, the actual behavior of the integral in which we are interested is $1/(q^2)^{\tau_{min}}$, where $\tau_{min}$ is 
the minimal twist, $i.e.$ the one corresponding to $l=l'=0$. One could also consider the vector meson case for generic (again even) $l$, which goes like $1/q^{l+4}$. 
In terms of $l'$ this reads $1/(q^2)^{l'+3-1}$. Define now $\tau=l'+3-1$, where the $-1$ stands for the fact that we have a spin 1 meson. 
Then the integral goes again like $1/(q^2)^{\tau_{min}}$. Note how in this case the different suppression factor $(r^2+L^2)$ in the integrand, is crucial to obtain 
the extra factor of $q$ which allows to interpret the exponent as a spin 1 hadron. Thus we see that the integral actually scales like $1/(q^2)^{\tau_{min}}$ 
for both the vector and scalar cases. The ``twist" is defined in terms of half of the $l$ corresponding to the actual meson 
state (the factor of 3 is related to the dimensionality of the ``basic" $l=0$ state, and it seems reasonable that it should be kept). 
This suggests the identification $\tau_{min}\leftrightarrow n_{min}/2$, where $n_{min}$ is the minimal number of valence partons in a QCD hadron 
(\textit{i.e.} 2 for a meson). Then upon using this dictionary, the form factor would scale like $1/q^2$, which is precisely the QCD result. 
However we must warn the reader that we do not have any compelling explanation for this identification.
 
This process has been recently considered \cite{Grigoryan:2008up} in the context of the hard wall model of \cite{Erlich:2005qh}, obtaining that the large $q^2$ behavior of the form factor matches that of QCD. However in that case the model is designed to capture the same symmetries as low energy QCD, so it is expected a good agreement.

\section{The complete unpolarized amplitude and inclusive processes}

It is interesting to compute the complete amplitude for the processes above. 
The physical process which we are actually looking at is really either $e\pi^0\rightarrow e\rho$ or $ef_0\rightarrow e\rho$ (or its crossed channel). 
Suppose we are interested in the unpolarized cross-section in figure (9).

\begin{figure}[!h]
\centering
\includegraphics[scale=0.5]{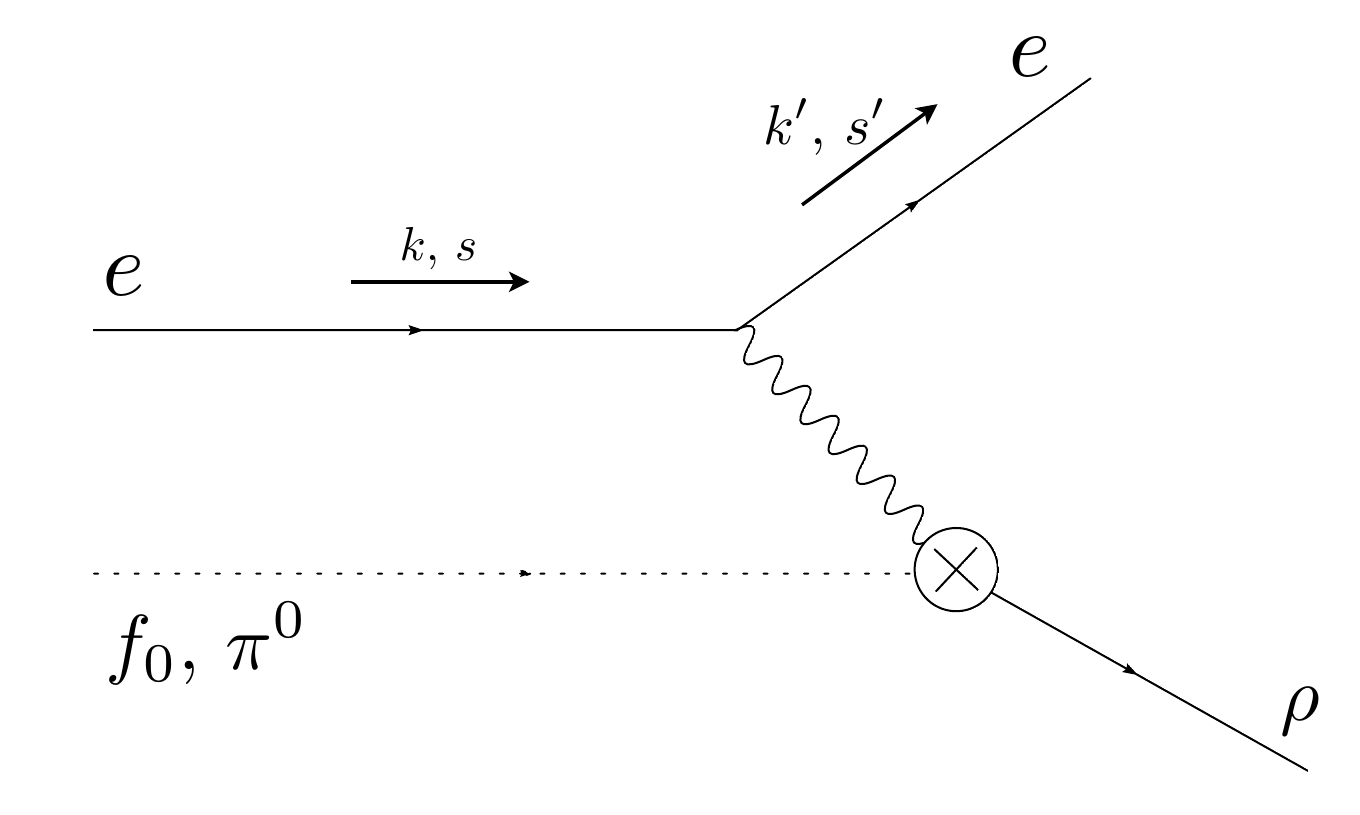}
\caption{Physical process.}
\end{figure}
The matrix element comes from 

\begin{equation}
i\mathcal{M}=-\frac{e^2}{q^2} \bar{u}_{s'}(k')\gamma^{\mu} u_s(k)\,\langle h_1|J_{\mu}|\rho\rangle\ .
\end{equation}
where $h_1$ stands for the initial hadron (either $f_0$ or $\pi^0$). Actually, the matrix elements $\langle h_1|J_{\mu}|\rho\rangle$ are nothing buth the ones we already computed.

Squaring, summing over polarizations and averaging over spins, this takes the usual form

\begin{equation}
|\mathcal{M}|^2=\frac{e^4}{q^4}L_{\mu\nu}W^{\mu\nu}\ ,\qquad W^{\mu\nu}=\sum_{pol.}|\langle h_1|J_{\mu}|\rho\rangle|^2\ .
\end{equation}
We can now compute $W^{\mu\nu}$ for each of the two cases in our theory. Interestingly in both cases the tensor structure leads, after summing over 
polarizations, to a Callan-Gross relation between the ``structure functions" (see appendix B). This should come as no surprise. 
We explicitly saw how the transverse character of our transitions was responsible for the large $q^2$ behavior of the form factors. 
The fact that we recover Callan-Gross scaling here is another consequence of having a transverse transition. \footnote{We thank G. Gabadadze for pointing this out to us.} 

Summarizing our results;

\vspace{0.5cm}
\textit{$f_0$ meson as in state:}
\vspace{0.2cm}

\begin{equation}
F_1^{n,m,l}=\frac{q^4}{4x^2}(F^{\sigma\rho}_{n,m,l})^2\ ;\qquad F_2^{n,m,l}=\frac{q^4}{2x}(F^{\sigma\rho}_{n,m,l})^2\ .
\end{equation}

\vspace{0.5cm}
\textit{$\pi^0$ meson as in state:}
\vspace{0.2cm}

\begin{equation}
F_1^{n,m,l}=\frac{q^4}{4x^2}\big(1+\frac{4x^2m_{\rho}^2}{q^2}\big)(F^{\pi\rho}_{n,m,l})^2\ ;\qquad F_2^{n,m,l}=\frac{q^4}{2x}(F^{\pi\rho}_{n,m,l})^2\ ;
\end{equation}
where we indicate the quantum numbers of the (pseudo)scalar $n$ and vector meson $m$, which share the same $l$. Note that the $x$ above is the Bjorken $x$, which in our quasielastic case is fixed eventhough we will keep it as open. 

The results above are quite reminiscent of DIS (Deep Inelastic Scattering) structure functions. 
Summing over possible final states we can construct an inelastic scattering amplitude. 
However we have to remember that our computation does not allow for the production of high spin states. 
Therefore if we want to interpret our results in terms of DIS we have to restrict ourselves to a regime in which the production of such states is highly suppressed. 
These high spin states are much more massive than the low spin states we have been considering in this paper. Therefore if we consider the 
DIS experiment in which we have a final state with $N$ particles (in our theory to leading order in $\lambda^{-1}$ $N=1$) whose 4-momenta add to $W$, whilst 
the initial hadron $h$ and off-shell photon have momenta respectively $p$ and $q$ - we have the trivial relation $W=p+q$. Squaring this we find

\begin{equation}
W^2-m_h^2=q^2(1-\frac{1}{x})\ .
\end{equation}

Since we don't want to allow for final state masses much larger than the initial state mass, we should only consider DIS in the region
where we take  $q^2\rightarrow -\infty$ and $x\rightarrow 1$ in such a way that $W^2-m_h^2$ remains finite and small. 
Thus we see that we could only access the $x\sim 1$ (\textit{i.e.} quasi-elastic) regime of DIS. In principle it should be possible to connect the 
threshold regime of the DIS with the form factors. At weak coupling this was studied in \cite{Drell:1969km, West:1970av}, however at strong coupling one must
be more careful since the calculation proceeds slightly differently \cite{Polchinski:2002jw}. 
Generically one would expect 

\begin{equation}
\label{guess}
F_2^{DIS}\sim F_2^{Form\,Factor} G(q^2(1-x^{-1}))\ .
\end{equation}
However a detailed discussion of these issues is beyond the scope of the current work.

\section{Conclusions}

In this paper we have been concerned with the structure of quark-antiquark bound states (mesons) at strong coupling. 
In order to study them we have probed these mesons with an external electromagnetic field. 
Together with the results \cite{Hong:2003jm, Polchinski:2001tt}, the picture that emerges is that the large $q^2$ behaviour of the matrix elements 
is dominated entirely by conformal invariance. 

This provides some justification as to why the scaling at weak coupling, based on a naive parton model, extrapolates to the strong coupling regime upon the 
replacement $n\leftrightarrow \tau$.
In the particular cases we have studied the helicity dependence of the matrix elements appears in a very explicit manner. 
The $U(1)$ form factors we computed are non-vanishing in the large $N$ limit, only for different in and out hadrons (\textit{i.e.} 
they are transition form factors). As we pointed out earlier, this is determined by the precise structure of the SUGRA lagrangian.
It also follows from the index structure and reality of the worldvolume fields on the brane, that the transitional form factors involve fields of different spin.
Therefore the helicity dependence of the matrix elements appears in an explicit manner.
The necessary powers of $q^2$, required to account for the helicity change in the amplitude, have a precise SUGRA origin in that they arise from terms
that are suppressed by additional powers of the warp factor.

We can imagine a way in which a different structure could arise. Refs. \cite{Erdmenger:2005bj} and \cite{Arean:2007nh} studied the Higgs phase of the theory,
in which the quarks have a non-trivial VEV and therefore the theory has a different vacuum structure.
In the gravity side this is achieved by means of a worldvolume instanton on the flavor branes (in order to go to the Higgs branch of the $\mathcal{N}=2$ theory one 
needs at least two flavors - see for example \cite{RodriguezGomez:2007za}). 
In the presence of the instanton, we could imagine new interaction operators emerging such as $\partial_{i}\Phi F_{inst}^{i r}F_{r \mu}^{A}\partial^{\mu}\Phi$ .
This term would capture a form factor for $\Phi$. Therefore studying the meson structure in the Higgs branch, and comparing it with what it was obtained in the 
Coulomb branch could be very interesting.

By using the same tools, we computed the $\gamma^*\gamma^*\pi$ form factor. However, in that case the comparison with the QCD result is more obscure. A priori it looks like our scaling is completely different ($1/q^6$ as oposed to $1/q^2$) from that in QCD. Alerted by the experience with the form factors, where the strong/weak coupling matching in the light of conformal invariance demands $\tau\leftrightarrow n$, we provided a first attempt of such a map by computing the form factor for generic $l$. Eventhough the matching requires some unjustifyied identification, we feel that it should be possible to have a deeper understanding of the particular scaling we obtained in the light of conformal invariance. We leave that issue open for future work. A consequence of gauge/gravity duality is that it satisfies VMD due to the (rather) generic properties of Sturm-Lioville operators. 
Since both the gauge field and the vector meson come from the same PDE, as described in appendix B, VMD follows directly.
This allows us to relate $F^{\gamma^*\pi}$ with $F^{\rho\pi}$ in very much of the same spirit as in QCD - where VMD also holds

There are a number of things which could be studied further. 
One is to understand the connection with inclusive processes (in particular DIS). It should be possible to compute DIS amplitudes directly in this model. 
One could consider computing the current-current correlator using a given hadron and non-normalizable mode wavefunctions, by employing the usual AdS/CFT methods. 
Understanding the Bjorken $x$ behavior of the DIS and computing the actual behavior of $G$ in (\ref{guess}) to compare with \cite{Polchinski:2002jw} 
would be very interesting and would surely shed more light on the structure of the mesons.

Another extension of our work could be to study the structure of the hadrons at high temperature. It has been suggested that these mesons could play an important role in the context of the QGP (see for instance \cite{Shuryak:2004cy} or \cite{Shuryak:2005yc} and references therein)
In the Minkowski phase of \cite{Mateos:2006nu} mesons still exist. It would be interesting to study the structure of the hadrons 
in that phase by probing them with photons. We expect the large $q^2$ behavior should not differ too much from the zero temperature result, 
since at $q^2\ll T^2$ one would expect to recover conformal invariance. 
However the IR behaviour will differ substantially. It would be interesting to check whether the relations between $F^{\rho\pi}$ and $F^{\sigma\rho}$ continue to be valid, 
and if any new form factors appear. 
A naive analysis in the light of our computations seems to suggest that there will be no modifications along these lines.

Finally we note that another way of getting vertices allowing for the same meson to appear as in and out state,  could be obtained by considering a 
pure gauge $B$-field along the Minkowski directions - in much the same spirit as in \cite{Filev:2007gb}. 
There are a number of things which could be studied this way, for example one could study the emission of photons by mesons in an external magnetic field. 
Interestingly in this case, even at zero mass the theory develops a condensate which breaks the $U(1)_R$ symmetry. 
It would be very interesting to check if one can reproduce more accurately the QCD $\gamma^*\gamma^*\pi$ behavior in this instance.

 \section*{Acknowledgments}

We are grateful to G. Gabadadze, A. Garcia-Garcia, C. Herzog, I. Klebanov, P. Kovtun, M. Papucci, A. Ramallo, A. Ritz, A.Scardicchio and H. Verlinde for useful discussions and comments.

D. R-G. acknowledges financial support from the European Commission through Marie Curie OIF grant contract no. MOIF-CT-2006-38381.

\appendix
\section{A brief review of VMD in gauge/gravity}

The hypothesis of vector meson dominance (VMD)  \cite{Sakurai:1960ju} regards the photon-hadron interaction at low energies 
in terms of an intermediate vector meson (see \cite{O'Connell:1995wf} for a review).

In our $D3$-$D7$ model, the mode dual to the EM current corresponds to the non-normalizable mode of the worldvolume $U(1)$ gauge field on the flavour brane with $l=0$. 
Restricting to the sector with $l=0$, both normalizable (\textit{i.e.} vector mesons with $l=0$) and non-normalizable (\textit{i.e.} EM current) modes 
come from solving the same PDE. This equation was first written down in \cite{Kruczenski:2003be}.   
Writing $\Psi=e^{iqx}\psi$, where  $\Psi$ stands for either normalizable or non-normalizable mode, it can be recast in terms of $w$ as follows

\begin{equation}
\partial_w\big(4w^2\partial_w\psi\big)-\frac{w}{1-w}\frac{R^4q^2}{L^2}\partial_{\mu}^2\psi=0\ ,
\end{equation}
For generic $q^2$ we would obtain the non-normalizable mode, whilst the normalizable mode appears when $q^2=m_{n,0}^2$. 
Let us now define

\begin{equation}
\mathcal{L}\psi=\partial_w\big(4w^2\partial_w\psi\big)\, ;\quad \lambda_q=\frac{R^4q^2}{L^2}\, ;\quad \rho=\frac{w}{1-w}\ ;\quad \tilde{\mathcal{L}}\psi=\mathcal{L}\psi-\lambda_q\rho\psi\ ;
\end{equation}
in such a way that the equation for the non-normalizable modes is just $\tilde{\mathcal{L}}A=j$ upon taking $j=0$. As usual we now write

\begin{equation}
j(w)=\int dw'\delta(w-w')j(w')\ ,\quad A=\int dw'G(w,w')j(w')\ ;
\end{equation}
where $G$ is the Green's function.

Clearly $\tilde{\mathcal{L}}$ satisfies Green's theorem, so

\begin{equation}
\label{GT}
\int_0^1dw\Big(\varphi\tilde{\mathcal{L}}\chi-\chi\tilde{\mathcal{L}}\varphi\Big)=4\Big(\varphi\partial_w\chi-\chi\partial_w\varphi)|_{w=1}\ ;
\end{equation}
where we have already eliminated the vanishing contribution from $w=0$. 
We can use (\ref{GT}) with $G$ and $A$, recalling that the non-normalizable mode satisfies 
Neumann boundary conditions at $w=1$ - and taking $j\rightarrow 0$ we can write the non-normalizable mode as

\begin{equation}
A(w,q^2)=\mathcal{N}\lim_{w'\rightarrow 1}\partial_{w'}G(w,w')\ ;
\end{equation}
where $\mathcal{N}$ is a constant. 
Thus we see that the non-normalizable mode is determined in terms of the Green's function of $\tilde{\mathcal{L}}$. 
In order to find an expression for $G$ we might consider the equation

\begin{equation}
\mathcal{L}\Phi_{n,0}^{II}+\lambda_n\rho\Phi_{n,0}^{II}=0\ ;
\end{equation}
where $\lambda_n=\lambda_q|_{q^2=m_{n,0}^2}$. This is nothing but the equation satisfied by the normalizable modes $\Phi_{n,0}^{II}$. 
We keep the subscript 0 to remind the reader that these normalizable modes correspond to vector fields with $l=0$. 
They satisfy the following completeness and orthonormality conditions

\begin{equation}
\sum_n\rho(w)\Phi_{n,0}^{II}(w)\Phi_{m,0}^{II}=\delta(w-w')\ ,\qquad\int_0^1dw\,\rho(w) \Phi_{n,0}^{II}(w)\Phi_{m,0}^{II}(w)=\delta_{nm}\ .
\end{equation}
Then if we consider

\begin{equation}
G=-\sum_n\frac{\Phi_{n,0}^{II}(w)\Phi_{n,0}^{II}(w')}{\lambda_n+\lambda_q}\ ,
\end{equation}
it is straightforward to check that this solves the Green's function equation for $\tilde{\mathcal{L}}$. Therefore the non-normalizable mode is given by

\begin{equation}
A(w,q^2)=\frac{m_q^2}{\lambda}\sum_n\frac{f_{n,0}\Phi^{II}_{n,0}(w)}{q^2+m_{n,0}^2}\ .
\end{equation}
and the decay constant of the $(n,0)$ vector meson is given by

\begin{equation}
f_{n,0}=\mathcal{N}\lim_{w'\rightarrow 1}\partial_{w'}\Phi^{II}_{n,0}(w')\ .
\end{equation}

\section{On the Callan-Gross relation}

We now turn to the appearance of the Callan-Gross relation result in a deeper way.  
In the context of VMD it was argued that this type of relation should naturally appear \cite{Sakurai:1969ss}. 
Also due to the structure of effective lagrangians we are considering, this  could be thought of as the low energy coupling of the Higgs to photons 
(see \cite{Shifman:1979eb, Kniehl:1995tn}). From that perspective one could argue that the effective vertex $hF^2$ involves, in particular, 
$h$ going into $\gamma\gamma$ through a top quark loop. Since the top quark is a spin $1/2$ particle for which one expects Callan-Gross, 
it seems reasonable to expect that this effective vertex also re-sums Callan-Gross.

\subsection{$F^{\mu\nu}F_{\mu\nu}$ contribution}

Consider 

\begin{equation}
\sum_{\epsilon} |(q_{\mu}\xi_{\nu}-q_{\nu}\xi_{\mu})(p^{\mu}\epsilon^{\nu}-p^{\nu}\epsilon^{\mu})|^2\ ,
\end{equation}
 where $\xi$ is the polarization vector of an external photon of momentum $q$ and $\epsilon$ the polarization of a massive vector particle of momentum $p$ and mass $M$. In addition $\xi\cdot q=0$, $\epsilon\cdot p=0$ and $-q^2=Q^2< 0$. This can be expanded as
 
 \begin{eqnarray}
&& 4\sum\Big((q\cdot p)^2|\xi\cdot \epsilon|^2-(q\cdot p) (\xi^*\cdot \epsilon^*)(q\cdot\epsilon)(p\cdot\xi)-(q\cdot\epsilon^*)(p\cdot\xi^*)(q\cdot p)(\xi\cdot\epsilon)+\nonumber \\ &&(q\cdot\epsilon^*)(p\cdot\xi^*)(q\cdot\epsilon)(p\cdot \xi)\Big)=\\ \nonumber &&\xi^*_{\mu}\xi_{\nu}\Big\{4\sum\Big((p\cdot q)^2(\epsilon^{\mu})^*\epsilon^{\nu}-(q\cdot p)\,p^{\mu}q_{\alpha}(\epsilon^{\alpha})^*\epsilon^{\nu}-(q\cdot p)\,p^{\nu}q_{\alpha}(\epsilon^{\mu})^*\epsilon^{\alpha}+p^{\mu}p^{\nu}q_{\alpha}q_{\beta}(\epsilon^{\alpha})^*\epsilon^{\beta}\Big)\Big\}
 \end{eqnarray}
 The sum runs over $\epsilon$ polarizations, so we have to use 
 
 \begin{equation}
 \sum_{\epsilon}(\epsilon^{\mu})^*\epsilon^{\nu}=(-\eta^{\mu\nu}+\frac{p^{\mu}p^{\nu}}{M^2})\ .
 \end{equation}
 After a little bit of algebra, one can show that the terms with $M$ cancel out, and one is left with
 
 \begin{equation}
4 \xi^*_{\mu}\xi_{\nu}\,\Big(-(p\cdot q)^2\eta^{\mu\nu}-q^2\,p^{\mu}p^{\nu}+(q\cdot p)\,(p^{\mu}q^{\nu}+p^{\nu}q^{\mu})\Big)\ .
\end{equation}
Consider now

\begin{equation}
(p\cdot q)^2\big(-\eta^{\mu\nu}+\frac{q^{\mu}q^{\nu}}{q^2}\big)-q^2\,(p^{\mu}+\frac{q^{\mu}}{2x})(p^{\nu}+\frac{q^{\nu}}{2x})\ ,
 \end{equation}
 where $x=\frac{-q^2}{2(p\cdot q)}$. After expanding this
 
 \begin{equation}
 -(p\cdot q)^2\eta^{\mu\nu}-q^2\, p^{\mu}p^{\nu}+(p\cdot q)\, (p^{\mu}q^{\nu}+p^{\nu}q^{\mu})\ ,
 \end{equation}
 so finally we have
 
 \begin{equation}
 4 \xi^*_{\mu}\xi_{\nu}\,\Big((p\cdot q)^2\big(-\eta^{\mu\nu}+\frac{q^{\mu}q^{\nu}}{q^2}\big)-q^2\,(p^{\mu}+\frac{q^{\mu}}{2x})(p^{\nu}+\frac{q^{\nu}}{2x})\Big)\ .
 \end{equation}
 This we can re-write as
 
  \begin{equation}
 4 \xi^*_{\mu}\xi_{\nu}\,\Big(\frac{4(p\cdot q)^2}{q^4}\frac{q^4}{4}\big(-\eta^{\mu\nu}+\frac{q^{\mu}q^{\nu}}{q^2}\big)+\frac{2x}{Q^2}\frac{q^4}{2x}\,(p^{\mu}+\frac{q^{\mu}}{2x})(p^{\nu}+\frac{q^{\nu}}{2x})\Big)\ .
 \end{equation}
where, as already illustrated, $-q^2=Q^2$. Then

 \begin{equation}
 \sum_{\epsilon} |(q_{\mu}\xi_{\nu}-q_{\nu}\xi_{\mu})(p^{\mu}\epsilon^{\nu}-p^{\nu}\epsilon^{\mu}|^2=4\xi_{\mu}\xi_{\nu}\hat{W}^{\mu\nu}\ ,
 \end{equation}
 with
 
 \begin{equation}
 \hat{W}^{\mu\nu}=\frac{q^4}{4x^2}\big(-\eta^{\mu\nu}+\frac{q^{\mu}q^{\nu}}{q^2}\big)+\frac{2x}{Q^2}\frac{q^4}{2x}\,(p^{\mu}+\frac{q^{\mu}}{2x})(p^{\nu}+\frac{q^{\nu}}{2x})\ .
\end{equation}
Therefore 

\begin{equation}
F_1=\frac{q^4}{4x^2}\ ,\qquad F_2=\frac{q^4}{2x}\ ;
\end{equation}
 so clearly $F_2=2xF_1$.
 
\subsection{$F_{\mu\nu}F_{\alpha\beta}\epsilon^{\mu\nu\alpha\beta}$ contribution}
 
 Consider now
 
 \begin{equation}
 \sum_{\epsilon} |(q_{\mu}\xi_{\nu}-q_{\nu}\xi_{\mu})(p_{\alpha}\epsilon_{\beta}-p_{\beta}\epsilon_{\alpha})\epsilon^{\alpha\beta\mu\nu}|^2
 \end{equation}
where again $\xi$ is the polarization vector of an external photon of momentum $q$, and $\epsilon$ is the polarization of a massive vector particle of 
momentum $p$ and mass $M$. In addition $\xi\cdot q=0$, $\epsilon\cdot p=0$  and $-q^2=Q^2> 0$. 
 
Since 
 
\begin{equation}
(q_{\mu}\xi_{\nu}-q_{\nu}\xi_{\mu})(p_{\alpha}\epsilon_{\beta}-p_{\beta}\epsilon_{\alpha})\epsilon^{\alpha\beta\mu\nu}=4q_{\mu}p_{\alpha}\xi_{\nu}\epsilon_{\beta}\epsilon^{\mu\nu\alpha\beta}\ ,
\end{equation}
we have, using the expression for the sum over polarizations

\begin{equation}
16q_{\mu}q_{\hat{\mu}}p_{\alpha}p_{\hat{\alpha}}\xi_{\nu}\xi^*_{\hat{\nu}}\epsilon^{\mu\nu\alpha\beta}\epsilon^{\hat{\mu}\hat{\nu}\hat{\alpha}\hat{\beta}}(-\eta_{\beta\hat{\beta}}+\frac{p_{\beta}p_{\hat{\beta}}}{M^2})\ .
\end{equation}
Therefore we get
 
 \begin{equation}
-16 \xi_{\mu}\xi^*_{\nu}\Big((q_{\rho}p_{\alpha}\epsilon^{\mu\rho\alpha\beta})(q_{\hat{\rho}}p_{\hat{\alpha}}\epsilon^{\nu\hat{\rho}\hat{\alpha}\hat{\beta}})\eta_{\beta\hat{\beta}}\Big)
 \end{equation}
 Making use of the properties of the $\epsilon$-tensor
 
 \begin{equation}
 (q_{\rho}p_{\alpha}\epsilon^{\mu\rho\alpha\beta})(q_{\hat{\rho}}p_{\hat{\alpha}}\epsilon^{\nu\hat{\rho}\hat{\alpha}\hat{\beta}})\eta_{\beta\hat{\beta}}=p^2q^2(-\eta^{\mu\nu}+\frac{q^{\mu}q^{\nu}}{q^2})+(p\cdot q)^2\eta^{\mu\nu}+q^2\,p^{\mu}p^{\nu}-(p\cdot)\,(p^{\mu}q^{\nu}+p^{\nu}q^{\mu})
 \end{equation}
 Adding and subtracting $(p\cdot q)^2\frac{q^{\mu}q^{\nu}}{q^2}$ we can re-write the expression above as

 \begin{equation}
 (q_{\rho}p_{\alpha}\epsilon^{\mu\rho\alpha\beta})(q_{\hat{\rho}}p_{\hat{\alpha}}\epsilon^{\nu\hat{\rho}\hat{\alpha}\hat{\beta}})\eta_{\beta\hat{\beta}}=(p^2q^2-(p\cdot q)^2)(-\eta^{\mu\nu}+\frac{q^{\mu}q^{\nu}}{q^2})+q^2\, (p^{\mu}+\frac{q^{\mu}}{2x})(p^{\nu}+\frac{q^{\nu}}{2x})\ .
 \end{equation}
 Since $p^2=M^2$ we can re-write this as
 
 \begin{equation}
 (q_{\rho}p_{\alpha}\epsilon^{\mu\rho\alpha\beta})(q_{\hat{\rho}}p_{\hat{\alpha}}\epsilon^{\nu\hat{\rho}\hat{\alpha}\hat{\beta}})\eta_{\beta\hat{\beta}}=-q^4(\frac{M^2}{q^2}+\frac{4(p\cdot q)^2}{4q^4})(-\eta^{\mu\nu}+\frac{q^{\mu}q^{\nu}}{q^2})-\frac{2x}{-q^2}\frac{q^4}{2x}q^2\, (p^{\mu}+\frac{q^{\mu}}{2x})(p^{\nu}+\frac{q^{\nu}}{2x})\ .
 \end{equation}
 Then
 
  \begin{equation}
 (q_{\rho}p_{\alpha}\epsilon^{\mu\rho\alpha\beta})(q_{\hat{\rho}}p_{\hat{\alpha}}\epsilon^{\nu\hat{\rho}\hat{\alpha}\hat{\beta}})\eta_{\beta\hat{\beta}}=-q^4(\frac{M^2}{q^2}+\frac{1}{4x^2})(-\eta^{\mu\nu}+\frac{q^{\mu}q^{\nu}}{q^2})-\frac{2x}{-q^2}\frac{q^4}{2x}q^2\, (p^{\mu}+\frac{q^{\mu}}{2x})(p^{\nu}+\frac{q^{\nu}}{2x})\ .
 \end{equation}

  \begin{equation}
 \sum_{\epsilon} |(q_{\mu}\xi_{\nu}-q_{\nu}\xi_{\mu})(p_{\alpha}\epsilon_{\beta}-p_{\beta}\epsilon_{\alpha})\epsilon^{\alpha\beta\mu\nu}|^2=16\xi^*_{\mu}\xi_{\nu}\hat{W}^{\mu\nu}\ ,
 \end{equation}
 with
 
 \begin{equation}
 \hat{W}^{\mu\nu}=q^4(\frac{1}{4x^2}-\frac{M^2}{Q^2})(-\eta^{\mu\nu}+\frac{q^{\mu}q^{\nu}}{q^2})+\frac{2x}{Q^2}\frac{q^4}{2x}q^2\, (p^{\mu}+\frac{q^{\mu}}{2x})(p^{\nu}+\frac{q^{\nu}}{2x})\ .
 \end{equation}
 In this case
 
 \begin{equation}
 F_1=q^4(\frac{1}{4x^2}-\frac{M^2}{Q^2})\ , \qquad F_2=\frac{q^4}{2x}\ .
 \end{equation}
 However we are working in a regime where $\frac{M^2}{Q^2}\ll 1$ whilst $x$ is fixed. Then $F_1\sim \frac{q^4}{2x}$, so we have that again
 
 \begin{equation}
 F_2=2xF_1\ .
 \end{equation}

\end{document}